# Imposing Regulation on Advanced Algorithms

**Fotios Fitsilis**

Head of Department for Scientific Documentation and Supervision

Scientific Service

Hellenic Parliament

Email: fitsilisf@parliament.gr
ORCID: https://orcid.org/0000-0003-1531-4128

This book would not have been possible without the support of my wife, especially in the later phase of her pregnancy! It is dedicated to her.

# Foreword

It is my very real pleasure to have the opportunity to write the foreword to this important new book by Dr. Fotios Fitsilis. The focus of his work is on how to regulate advanced algorithms, where the emergence of new technologies continues to present challenges for, among other things, science, law, and politics. For public lawyers, advanced algorithms raise particular questions about core aspects of constitutional and administrative law, including as to the nature and identity of algorithmic decision-makers, the manner in which decisions might be reviewed (and upon which grounds), and broader matters about accountability and control, not to mention conceptions of privacy. Such questions provide much of the backdrop to Dr. Fitsilis' book, which offers novel and insightful answers about some of the ways forward. I imagine that his commentary will become a key reference point for people working in this field.

Dr. Fitsilis' work focuses upon the role that two sets of actors might play in relation to advanced algorithms: regulatory bodies which operate within systems of multilevel governance; and Parliaments. He begins his analysis by examining how algorithms have evolved and how they now intersect with legal values such as non-discrimination, accountability and transparency. This provides the basis for a study of a number of key cases/legal moments involving algorithms, where he notes the approaches within a range of different jurisdictions and proposes a future role for a European algorithmic regulator. The remainder of his work considers the particular function that Parliaments might perform moving forward, where he argues that legislatures and executives need to work in tandem with technical and private-sector actors to set the parameters within which advanced algorithms might be developed. He here highlights the role that Parliamentary Research Services might play given their expertise and experience.

There are many reasons why Dr. Fitsilis' book might be regarded as important, though I would highlight two in particular. The first is that it represents a truly interdisciplinary approach to the topic at hand, as Dr. Fitsilis has a background in engineering and economics, as well as law. This is where much of the strength of his book is to be found, as the analysis within it borrows from his experience of working within and across those different academic disciplines. I would also note that he presently works within the Hellenic Parliament and that he thereby has a unique understanding of some of the issues that are analysed.

The second reason concerns the institutional setting in which many of the book's ideas were developed, namely the Academy of European Public Law in Greece. That Academy was founded in 1995 at a time when public law remained in a more traditional form, but its mission has since expanded given the challenges that public law now faces. Dr. Fitsilis' book stands as one of the most prominent examples of that expansion, and it is a testimony to his intellectual ability that he has been able to bring together his backgrounds in science and law. He is to be warmly congratulated on having done so in the form of this compelling and engaging book.

Gordon Anthony
Professor of Public Law
Queen's University, Belfast
Director of the Academy of European Public Law

28 June 2019

# Preface

Any end time scenario is a good reason to tackle the causing issue with scientific method and precision. This may very well be the case with Artificial Intelligence (AI). When discussing AI, most people think of so-called 'Artificial General Intelligence' (AGI), which refers, however, to the broader goal of reaching human-like intelligence, rather than to a single technology, or bunch of technologies. This goal is gradually approached through the inception of intelligent algorithms, i.e. special computer programs, rather than complex hardware, which are influencing an ever-expanding number of human activities. Within the course of this book, the notion of 'advanced algorithms' is developed to describe this kind of computer software, while calling for a timely regulation through well-defined, structured principles and dedicated government agencies with an oversight mandate.

My interest in the topic was sparked by my general research on the future of parliaments. Like every organization, parliaments will need to evolve to respond to digital challenges, otherwise they risk losing significance in the powers balance. Indeed, the concept of the 'smart' parliament involves the establishment of digital services for both internal and external (i.e. citizens) stakeholders. Among others, future parliaments will find themselves in the position of debating and examining the necessity to impose regulations to cover the use of advanced algorithms. Upon which criteria will the Members of Parliaments (MPs) decide? What kind of technical support and consultation would be necessary for them to reach educated, informed and sustainable law-making decisions?

While attempting to respond to such questions in some of my previous articles, it quickly became clear that the field of regulation of advanced algorithms needs to be widened to include both the *administrative* and the *judicial* levels. My understanding was further strengthened after attending a pair of workshops in 2017 and 2018, in Athens, Greece, under the name 'Artificial Cosmoi and the Law', organized, among others, by the Centre for Law, Economics and Society of University College London. This book as such builds upon a series of lectures on administrative law during the 2017 and 2018 sessions of the Academy of European Public Law, the essence of which was gradually developed and brought to the present form.

There were numerous challenges encountered whilst writing this book, mostly dealing with methodological issues as well as with the holistic approach that was apparently needed to discuss an omnipresent topic. Hence, a comparative legal perspective is not methodologically intended but necessary to shed light on the response of different legal orders to certain effects of advanced algorithms. In addition, there were inherent difficulties in choosing representative examples of algorithmic regulation from innumerous cases with significant national or international impact. There are just a few examples for regulating algorithms per dedicated legal provisions (the MiFID (EU) and the GDPR (EU) constitute exceptions to this rule and are therefore examined in detail later). The decision was therefore taken to select cases that were ruled on grounds of major areas of law, such as competi-tion, labor, environment, data-protection, consumer protection law and others. As a result, eight prominent case studies are presented in this book.

Back in my own field of study, the 'evolution of parliamentary systems and procedures', this book argues that parliamentarians and the Executive, supported by technical and private-sector consultations, should work closely together to define the legal grounds upon which advanced algorithms are to be developed and operated. In this respect, the role of Parliamentary Research Services (PaRS) and specialized agencies is of particular significance as they may provide a much needed in-depth analysis and strategic advice.

Overall, the book attempts a structured approach to systematize regulatory activities in the realm of advanced algorithms, a scientific field that is currently going through a development frenzy. Moreover, it constitutes a compact tool to provide guidance to regulators, be they lawmakers, judges or administrators. In this sense, as current bibliography is rather scarce, this book offers to all stakeholders a state-of-the-art introduction to the field of algorithmic regulation and supports the creation of guidelines for organizations on how to structure their regulatory activity.

## Acknowledgments

This book has one author but several supporters. Without their invaluable assistance, it would not have been possible to research and produce it in the form you are looking at.

In this regard, I would like to thank Prof. Georgios Yannopoulos, Law School of the National and Kapodistrian University of Athens, for his assistance in setting up a state-of-the-art structural framework for this book. Moreover, I would like to thank Prof. Spyridon Vlachopoulos, Law School of the National and Kapodistrian University of Athens, and Prof. Jean-Bernard Auby, Emeritus Public Law Professor of the Sciences Po, Paris, for providing legal-technical advice.

Special thanks go to Prof. Gordon Anthony, School of Law, Queen's University, Belfast, for the final review of the manuscript, for writing the foreword and for providing decisive comments. Many thanks also to Prof. Kostas Mavrias, Emeritus, Professor of Constitutional Law at the Law School of the National and Kapodistrian University of Athens and President of the Scientific Council of the Hellenic Parliament for his push towards scientific excellence and his steady support throughout the years. I would also like to underline the unseen contribution of many of my colleagues and fellow scientists, too numerous to mention, for their support and the fruitful discussions throughout the drafting phase.

Many thanks also to the friends and colleagues from the Hellenic Optical Character Recognition (OCR) Team, for their true dedication to science and innovation that has always been a source of inspiration to me. Dimitris Garantziotis deserves a special mention for his valuable contribution in the last phase of drafting. Lots of thanks go to Antonia Becou who professionally redesigned all graphic material in the book. Last, but not least, I would also like to thank Bruce Philip Todd, not only for the linguistic assessment and support, but also for the out-of-the-box general comments.

Above all, I would like to thank my family for all the support that gives me strength to carry on.

# Contents

**About the Author**

Dr. Fotios Fitsilis has over 20 years of professional experience in science positions within in both the private and the public sector. Since 2008, Dr. Fotios Fitsilis is Head of Department for Scientific Documentation and Supervision and lead researcher at the Scientific Service of the Hellenic Parliament. While operating on a global scale, he has been active in fields ranging from telecommunications and logistics to management and good governance, which has included recent papers on e-governance and institutional development including improvements to parliamentary oversight committees. Dr. Fitsilis has been visiting Professor for parliamentary procedures and legislative drafting at the Universidad Complutense de Madrid. In 2017, he founded the *Hellenic OCR Team*, a crowdsourcing initiative for the study of parliamentary data. Dr. Fitsilis has an academic background in Law (LL.M. in International Law), Economics (Diploma in Financial Engineering) and Engineering (Diploma in Electrical Engineering), while also holding a doctoral degree in Electrical Engineering.

## List of Abbreviations

| | |
|---|---|
| AEPL | Academy of European Public Law |
| AECD | Auxiliary Emission Control Device |
| AEDP | Agencia Española de Protección de Datos |
| AI | Artificial Intelligence |
| AGI | Artificial General Intelligence |
| Art. | Article |
| BGH | Bundesgerichtshof |
| CBPC | California Business and Professions Code |
| CEO | Chief Executive Officer |
| CJEU | Court of Justice of the European Union |
| CoC | Certificate of Conformity |
| COMPAS | Correctional Offender Management Profiling for Alternative Sanctions |
| DG | Direction Générale (European Commission) |
| DNN | Deep Neural Networks |
| DoJ | Department of Justice |
| DPA | Data Protection Agency |
| DPO | Data Protection Officer |
| EC | European Commission |
| ECHR | European Court of Human Rights |
| ECLI | European Case Law Identifier |
| ECT | European Communities Treaty |
| ENISA | European Union Agency for Network and Information Security |
| EP | European Parliament |
| EPA | Environmental Protection Agency |
| EPRS | European Parliament Research Service |
| ESMA | European Securities and Markets Authority |
| EU | European Union |
| EULA | End-User License Agreement |
| GDPR | General Data Protection Regulation |
| GUI | Graphical User Interface |
| HFAT | High-Frequency Algorithmic Trading |
| ICJ | International Court of Justice |
| ICO | Information Commissioner's Office (UK) |
| ICT | Information and Communications Technology |
| IE | Internet Explorer |
| IEEE | Institute of Electrical and Electronics Engineers |
| IFLA | International Federation of Library Associations and Institutions |
| IPU | Inter-Parliamentary Union |
| ISB | Independent State Body |
| ISP | Internet Service Provider |
| IT | Information Technology |
| JURI | European Parliament's Committee on Legal Affairs |
| MAR | Market Abuse Regulation |
| MEP | Member of the European Parliament |
| MiFID | Markets in Financial Instruments Directive |
| MiFIR | Markets in Financial Instruments Regulation |
| MP | Member of Parliament |
| NGO | Non-Governmental Organization |
| OBA | Online Behavioral Advertising |
| OCR | Optical Character Recognition |
| OTC | Over-The-Counter |
| PaRS | Parliamentary Research Services |
| PLS | Post-Legislative Scrutiny |
| PoLA | Principle of Least Astonishment |
| SABAM | Société belge des auteurs, compositeurs et éditeurs SCRL |
| SME | Small and Medium-sized Enterprise |
| TFEU | Treaty on the Functioning of the European Union |
| TGI | Tribunal de Grande Instance |
| UI | User Interface |
| UK | United Kingdom |
| URL | Uniform Resource Locator |
| US | United States |
| USC | United States Code |
| UWG | Gesetz gegen den unlauteren Wettbewerb |
| VW | Volkswagen |

## List of Figures

## List of Tables

# 1 Introduction, definitions and scope

**Fotios Fitsilis**

**Abstract**

This book focuses on cases where regulation has been imposed on advanced algorithms due to judicial or administrative decisions. From a series of different topics of algorithmic conduct, a number of case studies have been selected, in order to determine similarities or divergence in regulatory behavior. The development of regulatory bodies is therefore also discussed. Moreover, this book constitutes a first of its kind in terms of recording, classification and comparative legal assessment of significant cases of algorithmic regulation, with a view to establishing best practice and a responsible way forward.

Keywords: legal assessment, advanced algorithms, regulatory modes, algorithmic regulation, regulatory bodies.

For decades technological innovation was linked to the invention of *hardware* products. *Software*, that is computer programs, has been developed to drive hardware and little importance was given to it. The structured development of software as an engineering branch, as stated by the term 'software engineering', started much later.[1] However, it is not without irony that the sci-fi genre, rather than technologists, has first captured the strength and far-reaching consequences of algorithms, the cornerstone of software engineering, which, in popular culture, are usually paralleled with applications of Artificial General Intelligence (AGI).[2] One may just resemble the supercomputer Deep Thought in the novel *The Hitchhiker's Guide to the Galaxy,* from which originates the famous quote 'The answer to the ultimate question of life, the universe and everything is 42'.[3] Or HAL 9000 in *2001: A Space Odyssey*, an intelligent computer on board the spaceship Discovery, which malfunctions when presented with conflicting orders.[4] It is because of the rise of computer algorithms that large parts of our lives are now being managed by powerful applications of digital computing. This book is about regulation of algorithms. Moreover, it speaks about regulation of 'advanced algorithms'.

---

[1] On the history of software engineering see, e.g., Wirth (2008) and Booch (2018).

[2] In AGI, machine intelligence matches human intellectual properties; see, e.g., Goertzel and Pennachin (2007); the respective point in time this might take place is known as 'singularity'.

[3] See Adams (1979).

[4] See Clarke (1968).

Such algorithms are also influencing traditional institutions. Originating in Ancient Greece some 2.500 years ago, the concept of democracy is based on the core idea that governments are only legitimate when they are based on the consent of the people (the governed). In addition, according to Rousseau, citizens are both subject and sovereign, whereas law is the expression of the will of the people.[5] In contemporary democracies, people have to be guaranteed certain fundamental rights, such as the right to education, the right to equal protection before the law and the right to access to information.[6] However, what happens when the evolution of advanced algorithms, for instance in the form of digital communication platforms, is threatening these basic rights and, through that, the building elements of democracy?[7] Few would disagree that this situation calls for prompt regulatory measures, implementation of these measures and legislation, as well as oversight and enforcement capability.

As highlighted, the research field of the book is the regulation of advanced algorithms. It focuses on cases where regulation has been imposed on algorithms pursuant to judicial or administrative decisions. Its main objective is to identify possible common patterns behind these decisions[8] as guidance for future conduct with advanced algorithms. In the legal domain, the book examines the general question about the future of law in the era of advanced algorithms. Legal challenges presented by algorithms transcend different areas of law, which of course include administrative law.[9]

In order to analyze the state-of-play in the field, a thorough literature review has been conducted. In addition to already published research, several online resources by media, hackers, bloggers, investigative journalists and others have been used in this work. However, language barriers and issues of accessibility to case law limit this search to most 'advanced' legal orders. As a result, with a few exceptions regarding Japan and China, analysis is focused on relevant European and US case law. However, a comparative legal perspective is not methodologically intended but necessary to shed light on the response of different legal orders of selected states to certain effects of advanced algorithms. This discussion is of particular importance when we highlight the steps towards the development of regulatory bodies and their role, especially in the European context, where there are interesting questions to be asked about the overlap between global, supranational and national institutions and processes.

---

[5] On Rousseau's democratic principle and the surrounding discussion see, e.g., MacAdam (1983); the Universal Declaration of Human Rights states that '[t]he will of the people shall be the basis of the authority of government; (…)'; in United Nations (1948), Art. 21 §3; Cohen and Rogers (1983), p. 149, introduce the Principle of Democratic Legitimacy as a principle for public justification.

[6] United Nation's 1948 Universal Declaration of Human Rights constitutes a milestone document in the history of fundamental human rights; see also *supra* note 5.

[7] Sunstein (2017) illumines the dangers to democracy that arise out of the uncontrolled use of the internet.

[8] In the case of judicial decisions we generally speak of case law.

[9] A concise overview of this area of the law is provided by Leyland and Anthony (2012).

Concretely, the book attempts to answer two basic research questions:

1. Do advanced algorithms need to be regulated?
2. If so, who is going to be responsible for imposing such regulation?

While the answer to the first question is binary, i.e. 'yes' or 'no', there are a number of entities, that may impose algorithmic regulation. Several sources of regulation have been considered and are presented in this book. These may rely on a centralized or decentralized approach and may involve a dedicated regulatory body and/or even the parliament itself. Additional relevant issues, such as the degree of algorithmic regulation, have also been touched upon.

We publicly speak of algorithms and their regulation, but what exactly is an 'algorithm' and what is meant by the composite term 'algorithmic regulation'? Literature is not scarce when it comes to define an 'algorithm'. In a 1983 book on computer programming an algorithm is defined as:

> '(…) a finite sequence of instructions, each of which has a clear meaning and can be performed with a finite amount of effort in a finite length of time.'[10]

The etymology of the word is believed to be derived from al-Ḵwārizmī 'the man of Ḵwārizm', a corruption of the name given to the ninth century mathematician Abū Jaʿfar Muhammad ibn Mūsa, influenced by the Greek word *αριθμός* (arithmós i.e. number).[11] A common misunderstanding by non-computer scientists is that algorithms are restricted to computer-related processes, whereas in reality they are omnipresent and abundant in real life situations.[12] The personal morning routine or a cooking recipe are classic examples of real life algorithms. In general, an algorithm has the following attributes:

- Its steps are ordered,
- Its steps are well-defined,
- There is a limited number of steps and
- It produces a specific result.

---

[10] Definition taken from Aho et al. (1983), p. 2.

[11] According to Oxford Dictionaries, s.v. "algorithm", accessed May 30, 2019, https://en.oxforddictionaries.com/definition/algorithm.

[12] See, e.g., Louridas (2017).

In computer science, an algorithm can be understood as detailed instructions to a computer to perform a given task.[13] A computer program, plainly defined as a set of instructions to perform a certain task, may rely on one or more core algorithms.[14] Algorithms have been at the center of the development of Information and Communications Technologies (ICT). Moreover, their most advanced counterparts lie in the heart of potentially powerful as well as influential technologies, that drive social networks, trade in stock markets and calculate our taxes. Hence, in order to differentiate these more elaborated and complex algorithms from their less-developed predecessors, we adhere the term 'advanced algorithms' to describe them.[15] Fig. 1.1 shows the natural evolution of algorithms in four distinct steps, from simple tasks and processes that form some of the primary algorithms which are then embedded into more complex computer programs. The final step of evolution is currently populated by advanced algorithms. Following the technological singularity, this position will be taken over by AGI algorithms.

As algorithms evolve and increase in complexity an issue that needs to be dealt with is transparency in their operation. This is the so-called opacity problem and it is of particular importance when dealing with algorithmic decision-making systems.[16] In certain circumstances, their outcome may produce legal consequences that have implications for the legal rights of data subjects or, as we have seen in the case of digital communication platforms and social media, in the legitimacy of governance.[17]

When it comes to regulation –in this case, of digital platforms– few bring it closer to the point than Cédric O, France's Junior Minister for digital affairs:

> 'It is not for the platforms to decide what justice they should apply. The state makes the rules'.[18]

---

[13] Such instructions are encoded according to the typology of a computer language.

[14] Additional commands may define, among others, the User Interface (UI), data handling, communication protocols and interaction with other system parameters.

[15] However the term is not new and has been used in science for decades. For instance, Jayant (1986) and Wilamowski (2009) use the term to describe advanced coding techniques for voice communication and advanced learning algorithms for neural networks, respectively.

[16] See, e.g., Pasquale (2015) and Burrell (2016).

[17] In Chapter 2, examples of algorithmic opacity are going to be discussed within the discussion that touches upon basic principles of administrative law, such as non-discrimination and transparency.

[18] See relevant CNN Business report by Gold and Siad (2019); France pursues an increasingly aggressive strategy in the regulation of social networks, see Desmaris et al. (2019), whereas at the moment the European Commission (2018)

However, there is not a unified approach when defining the term 'regulation'. Orbach concludes that '[r]egulation is state intervention in the private domain, (…)'.[19] In the present context, the intervention of the state implies that its full arsenal of legislative and judicial tools may be employed to control, direct or manage the development and effects of application of advanced algorithms. Traditionally, command-and-control systems have been utilized to impose regulation. But in recent times, several new approaches have emerged.[20] Since we will be investigating different legal orders across several continents, one needs to bear in mind the paradox of calling for regulation in neo-liberal western democracies, where by definition deregulation should be the rule.[21] Certainly, there is always an option for States or supranational conglomerates not to proceed with any regulatory action.[22] The absence of regulation can be an option of choice, a product of ignorance or even mere inaction. However, we argue that, in cases of algorithmic conduct, no-regulation will inevitably lead to self-regulation in order for a natural equilibrium to be restored.[23]

We call these different approaches 'modes of regulation'. Several regulatory possibilities may exist within a mode. A classification of modes cannot be absolute, because it is not always possible to avoid overlapping between different ones.[24] Hence, indicatively, the essence of regulation may be captured within the following series of modes:[25]

1. Intervention mode:[26] command-and-control, self-regulation or co-regulation,
2. Hierarchical (or geographic) mode:[27] global, supranational, national (or local),

---

is relying on voluntary action by the stakeholders, through an EU Code of Practice on disinformation.

[19] In Orbach (2012), p. 10.

[20] See, e.g., Finck (2017) and Trubek and Trubek (2007).

[21] See Jordana and Levi-Faur (2004), p. 10.

[22] Gibbons (1997, p. 483) names no regulation of cyberspace a 'null choice'.

[23] See also Kleinsteuber (2004), p. 64.

[24] The timing mode, for instance, is event driven, i.e. regulation may be imposed before (*ex-ante*) or after (*ex-post*) a certain event has occurred; similarly, judicial regulation within the type mode is (mostly) triggered following a series of events that lead to legal action; in the following chapters, the type and the timing modes will frequently be used interchangeably; this does not mean that they are always identical, as legislative action may follow a judicial decision leading to a spiral of further actions from different stakeholders.

[25] More details about these modes, as well as a more narrowly-defined regulatory approach on advanced algorithms, will be presented in Chapter 4 and Chapter 5.

[26] Principles-based self-regulatory/coregulatory measures have been proposed by the European Commission to regulate online platforms; see, e.g., Finck (2017);

[27] This mode positions regulatory action in the chain of multi-level governance.

3. Natural mode:[28] direct or indirect,
4. Type mode:[29] legislative or judicial,
5. Timing mode:[30] *ex-ante* or *ex-post*.

During the research phase we have been presented with inherent difficulties in choosing representative examples of algorithmic regulation. Dozens of different cases of algorithmic conduct across several continents have been screened to find common ground for a detailed legal discussion. Some of these were promising candidates with potential necessity for algorithmic regulation. However, they were still developing stories with unclear endings.[31] Instead, the choice was made to select landmark cases where algorithmic regulation had already been imposed. But how to make choices from such a wide selection of cases? The decision was therefore made to select cases of algorithmic interest that were ruled on grounds of *major areas of law*.

Screening has also been performed at the legislative level. Not unsurprisingly, given the complexity of algorithmic regulation,[32] only a few examples of clear legislative action have been detected, such as Directive (EU) 2014/65 and Regulation (EU) 2016/679.[33] They constitute the European regulatory framework around algorithmic trading and data protection, respectively. However, it should be clear that these legal texts constitute exceptions to the rule and are therefore examined in detail.

As a result, a number of case studies of algorithmic regulation that touch upon eight major areas of law, such as competition, labor, environment, data-protection, consumer protection law and others, have been selected to be discussed in this book.[34] Each one of the six case studies with judicial intervention has been ruled based on one or more of these areas of law. Furthermore, the two mentioned legislative frameworks have been developed based on certain areas of law that lie at their core. In total, the eight areas of law that are displayed in Fig. 1.2 represent basic legal sectors that have been applied in cases of algorithmic regulation. We emphasize the fact that the examined case studies are not exclusively covered by the mentioned areas of law, nor that there is always a one-to-one match. For instance, the use of penal law for criminal investigations has been detected, but is not fully relevant for most of the cases.

---

[28] The nature of regulation is captured herein, i.e. whether regulation aims at changing the code of a given algorithm (direct regulation) or its environment, such as the behavior of its controller (indirect regulation).

[29] The type mode describes regulation coming out of two basic branches of government, legislative and judicial; bylaws and other administrative decisions are frequently not possible to be issued without prior legislative acts or judicial decisions.

[30] See *supra* note 24 and accompanying text.

[31] See, e.g., the COMPAS and the Amazon's Echo cases in Chapter 2.

[32] The number of possible regulatory modes, which were presented earlier, speaks for the complexity involved in imposing algorithmic regulation.

[33] This also goes under the name General Data Protection Regulation (GDPR).

[34] Six of the case studies represent legal disputes that were ruled before the courts (judicial regulation); in addition two legislative frameworks are discussed (GDPR and MiFID); these eight case studies are analyzed in Chapter 3. The effects of algorithms in the realm of administrative law are discussed in Chapter 2.

The topics of choice are specific cases of algorithmic regulation or stand-alone pieces of legislation that regulate a certain field, such as personal data protection.[35] In the Microsoft cases, the conditions that led to the separation of the operating system from the media player and the internet browser software are discussed. The Volkswagen emissions case dealt with an illegal algorithmic switch, which sensed the operating conditions of the vehicle and adjusted gas emissions accordingly. Ad-blocking was studied through a German case, *Axel Springer AG v. Eyeo*, which explored the inter-relations between online privacy, digital marketing and fair competition. The boundaries of the personal right to block online controversial content, widely known as the 'right to be forgotten', have been studied in the light of the case of *Google Spain SL and Google Inc. v. AEPD and González*. The area of 'sharing economy' has been approached through a series of high-profile cases against Airbnb and Uber. Finally, two topics have been dedicated to the discussion of algorithmic financial instruments, as regulated through Directive 2014/65/EU on markets in financial instruments, and personal data protection, as in the General Data Protection Regulation, 2016/679/EU. The main case attributes have been screened and compared, ranging from geographical location and administrative decisions to judicial reasoning and legal basis.

Apart from the above, the evolution of advanced algorithms is presented and a number of considerations are discussed, such as algorithmic bias, which may reproduce discriminatory behaviors as it is the case in real environments. The development of regulatory bodies is discussed in more detail as the complexity of advanced algorithms and their rapid evolution makes a traditional generalist approach rather inefficient. Within the same context, the role of parliaments is also discussed. Moreover, the cost and other general considerations of algorithmic regulation are tackled. The book also aims to systematize further study in the field of research. For this purpose, added-value is provided through data visualization by a series of tables and annexes. For example, Table A in the appendix presents the most significant court cases discussed herein, including complaints, decisions and other court documents. Table B in the appendix gives a list of laws, decisions, directives, regulations and resolutions from various national and international organizations, which have been analyzed in the course of this book.

The book, at its core, looks at universally applicable patterns in administrative decisions and judicial rulings. Analysis has been conducted to determine similarities or divergence in behavior among the different cases. Our assessment shows that in several of the cases presented, sources of general law, such as competition or labor law, are invoked as a legal basis, possibly due to the lack of additional specialized legislation on the subject area. In some occasions, it seems that a common law system is perhaps better placed to deal with this situation, as it can be more flexible. In a further step, the book investigates the role of regulatory bodies for advanced algorithms and considers the European Union Agency for Network and Information Security (ENISA), based in Heraklion, Crete, Greece, that focuses on network and information security, as an interesting candidate that could be tasked as the dedicated regulatory agency for advanced algorithms. A new EU Agency is not required and would not improve oversight or efficiency in this sector. The role of representative institutions in algorithmic regulation is also discussed. Today's parliaments may not yet be appropriately equipped, but their capacity can be strengthened to follow up on relevant regulatory provisions, e.g. in the context of Post-Legislative Scrutiny. The book concludes that despite the above concerns, governments should not be hesitant to invest in ameliorating the administrative state. Still, the relevant technologies are not ripe enough and we should use the time for the planning of regulatory principles and law-making. Scientific foresight and forward-thinking legal assessment should be widely employed in order to determine and regulate the effects of advanced algorithms in future societies. All in all, this book constitutes a first of its kind in terms of recording, classification and comparative legal assessment of significant cases of algorithmic regulation, with a view to establishing best practice and a responsible way forward.

---

[35] See Chapter 5 for an overview of analyzed topics and legal bases in the regulations of advanced algorithms.

Apart from this introductory part, the book is structured in four further chapters. Chapter 2 examines the evolution of algorithms and discusses broad concerns that are raised over their ethical utilization. The necessity of advanced algorithms and of their regulation is highlighted. In this context, the application of some of the core principles of administrative law, such as non-discrimination, accountability and transparency, is discussed. Chapter 3 is dedicated to landmark cases of algorithmic regulation. In total, eight cases of algorithmic conduct from several regulatory modes are discussed. Chapter 4 covers the development of regulatory bodies and the several forms of oversight institutions with regards to algorithms that have already been established. The role of representative institutions in algorithmic regulation is discussed and an existing EU agency, ENISA, is proposed as a potential candidate to take on the role of a pan-European regulatory body. Chapter 5 is devoted to the new perspectives around the regulation of advanced algorithms while presenting a summary of existing legal and administrative instruments which prove to be rather insufficient when it comes to confronting the array of issues and problems related to advanced algorithms.

# References

Adams, Douglas. 1979. *The Hitchhiker's Guide to the Galaxy*. London: Pan Books.

Aho, Alfred V., John E. Hopcroft, and Jeffrey D. Ullman. 1983. *Data structures and algorithms*. Boston: Addison-Wesley Longman Publishing Co.

Booch, Grady. 2018. The History of Software Engineering. *IEEE Software* 35(5): 108-114. https://doi.org/10.1109/MS.2018.3571234.

Burrell, Jenna. 2016. How the machine 'thinks': Understanding opacity in machine learning algorithms. *Big Data & Society* (June 2016): 1-12. https://doi.org/10.1177/2053951715622512.

Clarke C., Arthur. 1968. *2001: A Space Odyssey*. London: Hutchinson.

Cohen, Joshua and Joel Rogers. 1983. *On Democracy: Toward a transformation of American Society*. Middlesex: Penguin.

Desmaris, Sacha, Pierre Dubreuil, and Benoît Loutrel. 2019. Regulation of social networks – Facebook experiment. Interim mission report. May 2019. http://www.iicom.org/images/iic/themes/news/Reports/French-social-media-framework---May-2019.pdf. Accessed 7 June 2019.

European Commission. 2018. Code of Practice on Disinformation. https://ec.europa.eu/newsroom/dae/document.cfm?doc_id=54454. Accessed 6 June 2019.

Finck, Michèle. 2017. Digital Co-Regulation: Designing a Supranational Legal Framework for the Platform Economy. LSE Law, Society and Economy Working Papers 15/2017. http://dx.doi.org/10.2139/ssrn.2990043. Accessed 4 June 2019.

Gibbons, Llewellyn Joseph. 1997. No Regulation, Government Regulation, or Self-Regulation: Social Enforcement or Social Contracting for Governance in Cyberspace. *Cornell Journal of Law and Public Policy* 6(3): 475-551.

Goertzel, Ben, and Cassio Pennachin, eds. 2007. *Artificial General Intelligence*. Berlin, Heidelberg: Springer-Verlag.

Gold, Hadas, and Arnaud Siad. 2019. Why Mark Zuckerberg needed to impress Emmanuel Macron. *CNN Business*, May 10, 2019. https://edition.cnn.com/2019/05/10/tech/macron-zuckerberg-facebook-regulation/index.html. Accessed 5 June 2019.

Jayant, Nuggehally S. 1986. Coding speech at low bit rates: Advanced algorithms and hardware for voice telecommunications are paring hit rates by at least a factor of four, without losing intelligibility. *IEEE Spectrum* 23(8): 58-63. https://doi.org/10.1109/mspec.1986.6371061.

Jordana, Jacint, and David Levi-Faur, eds. 2004. *The Politics of Regulation: Institutions and Regulatory Reforms for the Age of Governance*. Cheltenham: Edward Elgar Publishing.

Kleinsteuber, Hans J. 2004. The Internet Between Regulation and Governance. In *The Media Freedom Internet Cookbook*, eds. Christian Möller and Arnaud Amouroux, 61-75. Vienna: OSCE.

Leyland, Peter, and Gordon Anthony. 2012. *Textbook on administrative law*. 7th ed. Oxford: Oxford University Press.

Louridas, Panos. 2017. *Real-world Algorithms: A Beginner's Guide*. Cambridge: MIT Press.

MacAdam, James. 1983. Rousseau's Democratic Principle. *Journal of the History of Philosophy* 21(2):231-234. https://doi.org/10.1353/hph.1983.0031.

Orbach, Barak. 2012. What Is Regulation? *Yale Journal on Regulation Online* 30(1): 1-10.

Pasquale, Frank. 2015. *Black Box Society*. Harvard: Harvard University Press.

Sunstein, Cass R. 2017. *#republic: Divided Democracy in the Age of Social Media*. Princeton: Princeton University Press.

Trubek, David M., and Louise G. Trubek. 2007. New Governance & Legal Regulation: Complementarity, Rivalry, and Transformation. *Columbia Journal of European Law* 13(3): 539-564.

United Nations. 1948. Universal Declaration of Human Rights. https://www.un.org/en/universal-declaration-human-rights. Accessed 11 June 2019.

Wilamowski, Bogdan M. 2009. Neural network architectures and learning algorithms. *IEEE Industrial Electronics Magazine* 3(4): 53-63. https://doi.org/10.1109/mie.2009.934790.

Wirth, Niklaus. 2008. A brief history of software engineering. *IEEE Annals of the History of Computing* 30(3): 32-39. https://doi.org/10.1109/MAHC.2008.33.

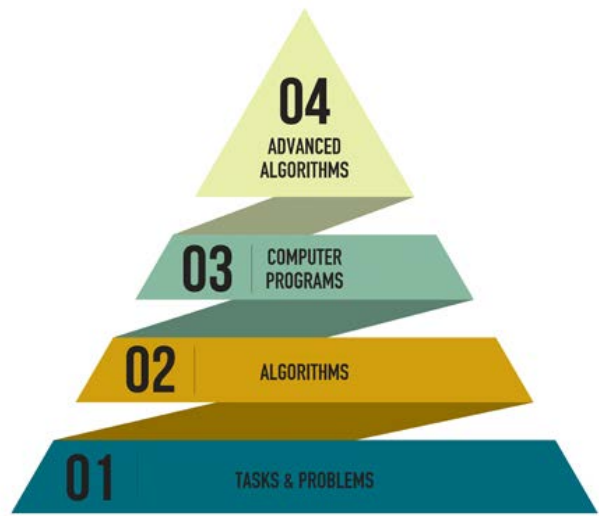

**Fig. 1.1** From simple tasks to advanced algorithms

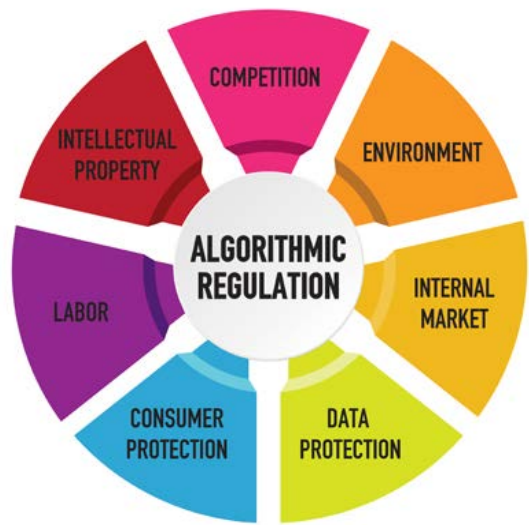

**Fig. 1.2** How algorithmic regulation touches upon major areas of law

# 2 Evolution of Advanced Algorithms

Fotios Fitsilis

**Abstract**

Technological evolution is based on advanced algorithms, but the pace of development raises concerns over their ethical, proper and legitimate utilization. This chapter discusses fundamental principles of administrative law, such as non-discrimination, accountability and transparency. As it seems, participation of advanced algorithms in the lives of millions is to a great extent irreversible and regulation will be needed in order to confront issues related to their development and valid operation.

Keywords: administrative law, algorithmic discrimination, accountability, transparency, counterfactual explanations.

## 2.1 Overview

New technologies that have the potential to re-shape human societies are emerging day by day in an unprecedented way and often at extraordinary and perhaps 'unhealthy' speed, as in the case of unregulated social media technology. Furthermore, it is difficult to find any aspect of our everyday lives that is not affected by these technologies, from commerce and farming, to medical care and education[1]. This technological revolution of our times is often characterized as the third technological revolution.[2] Gartner presents the top trends for emerging technologies in its annual 'Hype Cycle'.[3] From the broad range of fields presented therein, we extract the following as the most relevant to potentially affect law[4] and, in particular, *administrative law*: Artificial Intelligence (AI), Deep Learning, Machine Learn-ing, Cognitive Computing, Blockchain, Virtual Reality and Augmented Reality. These technologies have one thing in common; they are all heavily reliant on algorithms and, hence, we shall use in this book the collective term 'advanced algorithms' to describe them.[5] In the past, some of these technologies have been considered as being nothing more than mere science fiction. In the (not too distant) future, they may very well be the driving forces of human civilization.

---

[1] See, e.g., European Parliament (2017), p. 3.

[2] In Schwab (2016), Klaus Schwab, founder and executive chairman of the World Economic Forum, speaks of a 'fourth industrial revolution' that unlocks new opportunities in human-machine interaction.

[3] See Panetta (2017).

[4] Kemp (2016) presents legal aspects of three AI case studies, i.e. legal services, autonomous vehicles and smart contracts.

Advanced algorithms do not represent a technology of the future. Algorithms are already in force and here to stay. Their application in several economic and technical domains has already dramatically changed the way we do things. The ongoing automation of several industrial (and everyday) processes, previously performed by human beings, is saving a lot of time and resources while producing ever more accurate data and results. Notably, today, even warfare is changing compared to the past due to the use of drones and robotics, which allow for the engaging of targets through computer screens and dark rooms, thousand miles away from where the actual combat may be taking place.[6] This is why these technologies may also referred to as 'critical infrastructure'.[7]

The pace of development of these technologies raises legitimate ethical concerns over their proper utilization. Despite the overall undisputed benefits of advanced algorithms, their introduction in a range of sectors may be linked to certain dangers, or threats, that need to be addressed for these technologies to fully exploit their full and positive potential. The replacement of human labor by robots and algorithms in several levels of industrial production and administrative processes is causing a lot of anxiety around the rise of advanced algorithms. Taking this into consideration, societies and governments first need to clearly identify these concerns in order to proceed with effective oversight and control mechanisms, legislation and other countermeasures. The fact that man-made processes are now more accurate and fast, directly affects administrative science, as well as law. However, the application of advanced algorithms in the public domain can be problematic and may have several implications. Administrative law directly affects our everyday lives, from food security to public safety and security.[8] Thus, one of the purposes of this book is to examine from a holistic perspective the influence of these new technologies on administrative law. New technologies change today's societies in an unparalleled way and they will keep on transforming them. In bureaucratic procedures, we have managed to limit human errors extensively through the use of modern Information and Communication Technologies (ICT), including the internet and organizations' intranets. In public administration various organizational, managerial and archiving processes are usually redundant and time-consuming. Several of these are usually repetitive and non-critical tasks. The introduction of advanced algorithms can be decisive in the automation of these procedures, which may improve the quality of management and decision-making.[9] Hence, algorithms can be used to utilize employee output more efficiently, so that they spend quality working time on essential and qualitative tasks.

---

[5] See Chapter 1 for a description of the shift from 'algorithms' to 'advanced algorithms'.

[6] This is acknowledged in the relevant resolution by the Council of Europe (2015), as well as by Pasquale (2016).

[7] See Stone et al. (2016), p. 44.

[8] See Coglianese and Lehr (2017), p. 26.

The previous examples highlight the advantages in the use of advanced algorithms, but this does not mean that the introduction of such technologies comes without risks. In this regard, a first question that needs to be asked is to what extent we can legitimately and safely utilize algorithms in the administrative state, while maintaining an acceptable level of oversight and control? Certain fundamental principles of administrative law will be threatened if algorithms are left to operate without an appropriate legal framework.[10] In general, the principles of administrative law are concerned with human decisions involved in the exercise of state power and discretion. Furthermore, such principles offer a promising foundation for a regulatory framework for the growing number of algorithm-based decisions within the public sector. AI, as a sub-domain of advanced algorithms but maybe also the final frontier,[11] has attracted a lot of interest because of the aforementioned benefits. At the same time, it is being ferociously criticized for the possible negative outcomes of its application. This fear rises from the particular characteristics of algorithms. In the following section, some of the fundamental principles of administrative law, i.e. *non-discrimination*, *accountability* and *transparency*, will be discussed in the light of these emerging technologies.[12]

### 2.1.1 Non-discrimination

When it comes to law, many concerns are based on the fact that laws have been developed to apply to human beings, not to complex self-taught algorithms.[13] First of all, there are concerns regarding the discriminating 'behavior' of such systems, since the non-discrimination principle is an essential element of administrative law. Algorithms depend on variables and boundary conditions. Choosing the specific variables for an algorithm is not an objective task. It is the developer, the manager (private sector) or the state actor (public sector/government) who ultimately chooses the specific variables for a given scenario. This may lead to problems, or conflicts of interest, since these choices define the way the algorithm works, thus any bias in the variables may be directly translated within the technology into discriminatory results.[14]

---

[9] See Ng (2017).

[10] See European Parliament (2017), p. 6.

[11] In this case, we speak of Artificial General Intelligence; see, e.g., Goertzel and Pennachin (2007).

[12] When dealing with 'Explainable AI', a set of additional principles, each of which constitutes a separate research topic, may also apply, such as bias, fairness, transparency, safety, causality and engineering. More on the topic in Wierzynski (2018).

[13] See Coglianese and Lehr (2017), p. 6.

Moreover, there is a well-founded concern that inherently discriminatory automated processes will prevail in the decision-making because of the evolutionary nature of some algorithms, which is where establishing an acceptable level of control comes in. However, such concerns are not only known within the context of advanced algorithms. Bias and partisanship are also part of human nature and behavior.[15] Humans often succumb to discriminatory practices. Hence, when it comes to the incorporation of advanced algorithms into administrative processes, discrimination must be avoided, independently of who makes the decision, be it a human or an algorithm.[16]

At this point, it needs to be noted that even if algorithms engender biased outcomes, this does not necessarily mean that such outcomes are, or should be regarded, as legally problematic. It is possible for administrative decision-makers to differentiate between applications made to them, so long as any differential treatment is justified and proportionate.[17]

### 2.1.2 Accountability

Questions of accountability become even more urgent in relation to the use of algorithms.[18] In the traditional context of the administrative state, it is the people who make specific decisions which naturally have certain (legal) consequences. Thus, they can be held accountable for their actions or omissions. It has been fairly easy to locate the exact person(s) who have been responsible for a specific administrative action and to hold them accountable for it.[19] Unfortunately, that may not be the case with some algorithms due to the ambiguous nature of the decision-making processes. Algorithms are creations of the human intellect, which obey their core code. Similarly, in the past, it has been possible to track down the person responsible for a programming fault, e.g. for conducting wrongful/illegal bank transactions or influencing an industrial facility. However, advanced algorithms incorporate deep-learning, self-learning, fuzzy logic and a number of advanced technologies that largely decouple the human developer from the creation, that of the algorithm. When utilizing such technology, mistakes or mishaps will happen, just like in the case of their human operatives. In such cases, who (or what) is to be held accountable and which are the (legal) consequences? Ultimately, in the administrative state, does it make any (legal) sense to hold an algorithm accountable for a wrong or bad decision? In the criminal law domain, particularly in US, where great importance to the concept of *mens rea*, the intending mind, is attached, challenges are raised when advanced algorithms come into play.[20] Clearly, it is at best questionable to place accountability on machines and complex codes than on the humans who realized the manufacture and programming of the machines or developed the code.

---

[14] See, e.g., Ng (2017); Williams et al. (2018) assesses 'unacceptable judgement(s)' of algorithmic decision-making processes.

[15] See, e.g., Ng (2017).

[16] See Coglianese and Lehr (2017), p. 59; Algorithmic bias is further analyzed in *infra* Section 2.2.

[17] This point is also true, to a certain extent, when discussing issues of 'accountability' and 'transparency', though there might be an argument that those are core values that underlie the principles of administrative law.

[18] See, e.g., Ng (2017).

[19] On the legal nature of (automated) administrative actions see, e.g., Lazaratos (1989).

A recent direction to answering this legal dilemma has been provided in 2016 in response to the so-called 'monkey selfie case', where copyrights have not been provided to a monkey that shot a 'selfie'.[21] Following this decision, the US Copyright Office updated its rules which now include a passage stating that 'works produced by a machine or mere mechanical process that operates randomly or automatically without any creative input or intervention from a human author' do not qualify for copyright protection under US law. However, there is legal dispute whether the above decision tackles issues of 'ownership'. Similarly, under US Patent Law, there is not an explicit prohibition of protection for inventions of AI.[22]

Where there is no legal ground, and in the instance of AI, where this is the case, one may resort to alternative readings to look for guidance. For once, there are the Asimov's laws, also known as the 'Three Laws of Robotics' (a fourth one was added later), which aim to protect the integrity of human lives when interacting with robots. Even though these laws have been created for science fiction novels, they are considered crucial in modern discourse concerning AI and robots. In this regard, they have also been considered in the European Parliament's report with recommendations to the Commission on Civil Law Rules on Robotics.[23] In the light of the above, legislators will need to take interdisciplinary action in order to examine ways to at least partially regulate this highly dynamic scientific field.

---

[20] See Stone et al. (2016), p. 47.
[21] *Naruto v. David John Slater et al.*, 3:2015-cv-04324 (9th Cir.).
[22] Title 35 of the U.S. Patent Code.
[23] See European Parliament (2017), p. 6.

### *2.1.3 Transparency*

As discussed above, it is difficult to assess whether an algorithm is truly accountable, due to its so-called 'black box' nature, which prohibits the deeper understanding of how and why it reached a conclusion in a particular task.[24] This means that the reasoning behind its decisions may not be evident. If it is not clear how an algorithm has come to a decision, it may be particularly difficult to detect any bias in the decision-making process and to test if a decision is fair or not. This potential lack of transparency needs to be addressed carefully in an administrative state, since there may be serious concerns over the safeguard of democratic governance with the introduction of AI.[25] In addition, lack of transparency, which is an essential value of the democratic administrative state, due to the use of advanced algorithms, may lead to the questioning of legitimacy of administrative decisions.

The European Union is attempting to tackle the 'black box' problem and the resulting transparency issue by providing to EU citizens the so-called 'right to explanation'.[26] The 'right to explanation' can be identical with the personal right to be provided with 'meaningful information about the logic involved' in automated decisions.[27] However, scholars argue on the legal existence and feasibility of such a theoretical right.[28] In principle, the 'right to explanation' intersects with the duty to give reasons for decisions, be it administrative[29] or judicial.[30] Moreover, it relies on the well-founded right to information.[31]

Another area of concern is related to the impact of advanced algorithms on employment. Replacing humans with machines has been a major trend since the first industrial revolution, a trend that led to the formation of well-organized unions, as a response to the threat to job security. Algorithms and machines are therefore trusted, to an extent, to be faster, stronger and more efficient and effective than human beings. The term 'digital unemployment' has recently been coined to reflect the loss of jobs because of the rise of digital technology. Currently, only standardized procedures have been taken over by algorithms, while more complex tasks remain, at least for now, in human hands. In the past, machines have been seen as a way to 'liberate' the human race from hard labor, but today they may pose a threat to a multitude of professions, both blue collar and white collar ones, even to that of the judge. While it may be disputed whether an AI would be ever able to completely replace a human judge, this may very well be the case in standardized cases in lower instances of administrative courts. Recent research using Natural Language Processing and Machine Learning technology was able to establish predictive models in judicial decisions.[32] Nevertheless, new studies are projecting AI under a whole new light, suggesting that its inherent dynamics regarding employment might have changed, and are projecting two million new jobs in 2025.[33]

---

[24] See Coglianese and Lehr (2017), p. 18.

[25] See Coglianese and Lehr (2017), p. 49.

[26] This is attempted via the General Data Protection Regulation (EU) 2016/679, though there is no single provision therein labeled as such.

[27] According to Selbst and Powles (2017), p. 242.

[28] See, e.g., Wachter et al. (2017).

[29] In the case of administrative decisions in international law, a legal duty for administrators appears to be contentious; see Hepburn (2012), p. 641.

[30] On the justification of judicial decisions see Shuman (1971).

[31] See, e.g., Peled and Rabin (2011).

Another, positive this time, influence of new technologies is related to the introduction of ICT in the judicial sector, as part of accepted and necessary reforms across both the civil and criminal justice sectors. Currently, a large effort to network judicial institutions is widely materializing, thus making judicial decisions easily accessible and the judicial process significantly more efficient.[34] This development towards an inter-linked judicial community is expected to continue with the standardization of judicial documents, e.g. using the Akoma Ntoso XML-schema for legal documents.[35] When it comes to legal informatics, it has been shown that discrepancies in the legal outcome may arise when digitizing formerly manual procedures.[36] Given that technology (the code) is a regulatory modality affecting human behavior, it should be ensured that the law is the driving force behind it. To achieve this goal, 'the content of a law-related system should be defined, having as a guideline the hierarchy of legal sources prescribed by the legal order' and then be tested on whether it serves its purpose satisfactory.[37]

There are many additional challenges to be dealt with when handling new technologies, including, but not limited to, the increase in existing inequalities, the breach or invasion of private life, or the enhancement of discriminatory practices.[38] These challenges exist not only for governments but also for the private sec-tor, which is pioneering the field. However, one should keep in mind that only governments enjoy over the regulatory initiative and the means to enforce it.

---

[32] See relevant research by Aletras et al. (2016).

[33] See Gartner (2017).

[34] The Council of the European Union in 2010 has taken the decision to implement the European Case Law Identifier (ECLI), which is a human readable and computer processable code that be assigned to every judicial decision from every national or European court in order to facilitate cross-border accessibility of case law; see also the online resources of BO ECLI (2017), an EU project involving several partners from 10 member states.

[35] In 2018, Akoma Ntoso has become an OASIS standard; the document architecture, the relevant schemes and the general framework for legal documents are available at Akoma Ntoso (2018); detailed information on the rationale of standard-based management of legal documents as well as a discussion of semantic resources is presented by Sartor et al. (2011).

[36] See Yannopoulos (2012), p. 1024.

[37] See Yannopoulos (2012), p. 1028.

[38] According to Stone et al. (2016).

## 2.2 Necessity of Advanced Algorithms and their Regulation

The discussion about the necessity of advanced algorithms in contemporary and future societies, as well as the parameters of their regulation, is a complex one. A basic set of research questions has been posed in the introductory chapter. In order to approach it in a structured order, it can be further broken down into a more specialized set of sub-questions, such as the following:

- Is the use of advanced algorithms inevitable for open societies that need to regulate the lives of millions or even billions of people?
- Which are the underlying ethical dilemmas when developing or applying advanced algorithms?

It is outside the context of this book to tackle all these issues in detail. Instead, a more pragmatic approach is followed in order to discuss the relevant framework by presenting a series of recent topics related to advanced algorithms. At the same time, the book sheds light on the regulatory principles or guidelines that state actors, be it the government or the judiciary, pursue in cases that involve utilization of advanced algorithms. Hence, we will try to support earlier claims that indicate shortcomings in existing legal instruments to confront issues related to advanced algorithms.[39]

Algorithms are omnipresent. In some areas, such as the case of welfare delivery and monitoring, their integration is progressing 'at a breathtaking pace with little or no political discussion about their impacts'.[40] Yet, the inherent properties of software that make it attractive for certain applications can be potentially dangerous for others.[41] Advanced algorithms in the form of always-on smart home devises and digital assistants[42] have entered into the mainstream consumer market and enjoy a wide acceptance despite well-founded criticism. That such devices may pose a threat to privacy has become visible in the US case of an Amazon's Echo device, which has recorded and transmitted, without prior consent, a private com-munication to a random third party.[43] Earlier, in 2017, it was demonstrated that such a device could be turned into a wiretap.[44] While it is unclear whether there will be an investigation from the competent authorities, or any legal action, it becomes clear that the use of such devises is linked to serious privacy and security threats that have neither been tackled, nor fully understood as of yet. In a further case, a matter of, literally, life and death, dramatically increased public interest and awareness in the UK to the potentially negative outcomes of applications of advanced algorithms.

---

[39] See Barocas (2014).

[40] See Eubanks (2018), pp. 11-12.

[41] According to Grimmelmann (2005, p. 1758); in this regard, it is discussed whether software can be a good regulatory tool for rule- and standard-based decisions. This distinction is considered a staple of jurisprudential theory, since software applies rules rather than standards (Grimmelmann 2005, p. 1732).

[42] See, e.g., Google assistant (2018), a digital assistant that may handle a variety of everyday tasks up to a full first-person conversation with a human counterpart.

In May 2018, it was reported that a bug in the information system of the National Health Service in Britain may have caused, from 2009 to date, the premature death of up to 270 women suffering from cancer.[45] An algorithm that prevented sending alerts for planned mammograms to a specific age group of women is considered 'responsible' for this mishap. However, there is still much controversy among experts about how this 'algorithm failure' came about and whether or not any scandal had happened at all.[46] A further concern in relation to advanced algorithms is their sometimes biased operation. Although one would expect such algorithms not to express favoritism, these may rely on existing data sets from real environments or they may depict the mindset of their human programmers. Hence, if no proper or effective counteraction is taken, advanced algorithms may simply propagate or even maximize the original bias. A source of further consideration is that the most vulnerable parts of society are the ones that are going to be more prone to being exposed to algorithmic decisions and their potentially discriminatory outcomes.[47] This is a notion that seems to be attracting an increasing level of support.[48]

As computer programs and their underlying electronic systems are becoming more complex and opaque,[49] the task to debug advanced algorithms may prove challenging. Such a discriminatory behavior may be either intentional or unintentional. An example of an intentionally induced discriminatory behavior is differential pricing.[50] The Correctional Offender Management Profiling for Alternative

---

[43] According to a KIRO7 report by Horcher (2018).

[44] See Barnes (2017).

[45] Jeremy Hunt, Secretary of State for Health and Social Care, brought the relevant issue to the House of Commons on 2 May 2018. The statement and the consequent debate may be found in the minutes of proceedings of the House; see House of Commons (2018).

[46] Charette (2018) presents the public debate in an IEEE Spectrum report.

[47] See, e.g., O'Neil (2016).

[48] See also Eubanks (2018).

[49] There exists rich literature on the problems that arise due to the opacity of advanced algorithms. For more information, the reader may refer to Pascuale (2015), Burrell (2016) and de Laat (2017).

[50] The term is used when online users get different prices for the same product, depending on the result of a system evaluation of their available personal data, al-

Sanctions (COMPAS) algorithm offers one of the most prominent examples of un­intentional discriminatory behavior. This is a computer program developed by Northpointe, Inc. in US, in order to provide judges with risk assessments for the future criminal behavior of convicted individuals at every stage of the US criminal justice system. The US Department of Justice supports the use of such tools and a relevant sentencing reform bill is currently pending in Congress.[51] Previous evalu­ations of the COMPAS risk models were encouraging[52] and assessed that the algo­rithm '(…) performs well in predicting risk for offenders released from jail pretri­al'.[53] However, in 2016, a larger independent examination by ProPublica[54] showed that the algorithm was in fact biased against black prisoners. The company con­tested the results of this study but declined to provide access to the algorithm, which calculates a risk assessment score based on 137 questions either answered by offenders or retrieved from their criminal records.[55]

In a controversial case of prison labor, inmates in Finland have been used by a startup company to classify data used in the training of AI algorithms[56]. The latter argues that this is a win-win situation for both the company and the prisoners, as the approach helps the prisoners develop modern work skills. However, this thesis does not remain uncontested both ethically, as well as in its essence.[57]

Cases such as these have long been sources of concern, particularly among hu­man rights activists and data ethics scientists. With advances in machine learning technology, these voices are becoming louder, even predicting that 'AI could spell the end for the human race'.[58]

All these separate cases, whether small-scale incidents, as in Amazon's Echo case, or serious systemic errors, as in the COMPAS case, demonstrate that these concerns have a solid foundation and need to be tackled sooner, rather than later. Recently, ICT giants have taken significant steps towards a more people-centric

---

so called 'profiling'. Consequently, a used assessed as 'wealthy' may get a higher price compared to the average user. According to Regulation (EU) 2016/679 –this is the General Data Protection Regulation– profiling is only allowed following ex­plicit user consent.

[51] This is the Sentencing Reform and Corrections Act of 2015 at the 114th Congress (2015-2016).

[52] See concluding remarks in Brennan et al. (2009), p. 34.

[53] See Blomberg et al. (2010), p. 91.

[54] ProPublica is an independent, nonprofit network of investigative journalists.

[55] See the relevant report by Angwin et al. (2016).

[56] The case was presented by Chen (2019) in theverge.com.

[57] By Sarah T. Roberts, University of California, Los Angeles; in Chen (2019), *supra* note 56.

[58] According to Hawking (2014).

approach to AI.[59] Researchers are currently developing tools to help understand the internal operation of (deep) machine learning systems, a research field that is called 'Explainable AI'.[60] IBM has announced an open-source toolkit to check for and mitigate unwanted bias in datasets, machine learning models and state-of-the-art algorithms.[61] Another relatively new concept is the previously mentioned 'right to explanation'.

Should such a right be recognized to legal subjects, it could cause developers to structure their algorithms and systems differently in order to offer a higher degree of transparency. This could seriously affect their functionality and operation. Moreover, it is not always possible for the general public nor the developer to understand how an automated system has reached a particular decision. To make things worse, automated decision-making systems, e.g. in the form of Deep Neural Networks (DNNs), are changing rapidly because they constantly learn. As a result, an explanation that applies when at a certain stage of a DNN may not be the case in another. This is why a minimal solution has been developed in the form of what is called 'counterfactual explanations'. Counterfactual explanations 'provide insight into which external facts could be different in order to arrive at a desired outcome'.[62] Other scholars call for paying more attention to socio-technical systems, in order to come up with suggestions for the improvement of algorithmic regulation.[63] From the above, it can be deducted that participation of advanced algorithms in the lives of millions is irreversible to a great extent. Regulation is needed and several reasons that speak for such regulation have been discussed, such as biased operation and algorithmic malpractice.

Chapter 3 presents in more detail a number of topics of algorithmic regulation. However, the extent of regulation will inevitably vary from case to case. Societies (rather than functionaries) will have to determine the degree of regulation and this is why the role of representative institutions is eventually going to be important.[64]

## References

Akoma Ntoso. 2018. Version 1.0 Part 1: XML Vocabulary. OASIS Standard. Edited by Monica Palmirani, Roger Sperberg, Grant Vergottini, and Fabio Vitali. http://docs.oasis-

---

[59] See for instance Google's People + AI Research program, a 'Human-centered research and design to make AI partnerships productive, enjoyable, and fair', according to Google PAIR (2018).

[60] See, e.g., Wierzynski (2018).

[61] See Varshney (2018).

[62] See Wachter et al. (2018), p. 880.

[63] See Medina (2015).

[64] The role of parliaments in the regulatory framework of advanced algorithms is discussed in Chapter 4: Development of Regulatory Bodies

open.org/legaldocml/akn-core/v1.0/os/part1-vocabulary/akn-core-v1.0-os-part1-vocabulary.html. Accessed 5 June 2019.

Aletras, Nikolaos, Dimitrios Tsarapatsanis, Daniel Preoțiuc-Pietro, and Vasileios Lampos. 2016. Predicting judicial decisions of the European Court of Human Rights: a Natural Language Processing perspective. *PeerJ Computer Science* 2:e93. https://doi.org/10.7717/peerj-cs.93.

Angwin, Julia, Jeff Larson, Surya Mattu, and Lauren Kirchner. 2016. Machine bias. *ProPublica*, May 23. https://www.propublica.org/article/machine-bias-risk-assessments-in-criminal-sentencing. Accessed on 7 July 2018.

Barnes, Mark. 2017. Alexa, are you listening? MRW Infosecurity. August 1. https://labs.mwrinfosecurity.com/blog/alexa-are-you-listening. Accessed 22 March 2019.

Barocas, Solon. 2014. Data mining and the discourse on discrimination. Conference on Knowledge Discovery and Data Mining. August 24-27. New York City. https://dataethics.github.io/proceedings/DataMiningandtheDiscourseOnDiscrimination.pdf. Accessed 22 March 2019.

Blomberg, Thomas, William Bales, Karen Mann, Ryan Meldrum, and Joe Nedelec. 2010. Validation of the COMPAS risk assessment classification instrument. http://criminology.fsu.edu/wp-content/uploads/Validation-of-the-COMPAS-Risk-Assessment-Classification-Instrument.pdf. Accessed 28 June 2018.

BO ECLI. 2017. Building on European Case Law Identifier. http://bo-ecli.eu/. Accessed 22 March 2019.

Brennan, Tim, William Dieterich, and Beate Ehret. 2009. Evaluating the predictive validity of the COMPAS risk and needs assessment system. *Criminal Justice and Behavior* 36 (1): 21-40. https://doi.org/10.1177/0093854808326545.

Burrell, Jenna. 2016. How the machine 'thinks': Understanding opacity in machine learning algorithms. Big Data & Society (June 2016): 1-12. https://doi.org/10.1177/2053951715622512.

Charette, Robert. N.. 2018. 450,000 Women Missed Breast Cancer Screenings Due to "Algorithm Failure". *IEEE Spectrum*, May 11. https://spectrum.ieee.org/riskfactor/computing/it/450000-woman-missed-breast-cancer-screening-exams-in-uk-due-to-algorithm-failure.

Chen, Angela. 2019. Inmates in Finland are training AI as part of prison labor. *TheVerge.com*, March 28. https://www.theverge.com/2019/3/28/18285572/prison-labor-finland-artificial-intelligence-data-tagging-vainu. Accessed 5 June 2019.

Coglianese, Cary, and David Lehr. 2017. Regulating by Robot: Administrative Decision Making in the Machine-Learning Era. Research Paper No. 17-8. Institute for Law and Economics. University of Pennsylvania. https://www.law.upenn.edu/live/files/6329-coglianese-and-lehr-regulating-by-robot-penn-ile. Accessed 22 March 2019.

Council of Europe. 2015. Drones and targeted killings: the need to uphold human rights and international law. Resolution 2051. http://assembly.coe.int/nw/xml/XRef/Xref-DocDetails-EN.asp?FileID=21746&lang=EN. Accessed 22 March 2019.

de Laat, Paul B. 2017. Algorithmic Decision-Making Based on Machine Learning from Big Data: Can Transparency Restore Accountability? *Philosophy & Technology* 31: 525–541. https://doi.org/10.1007/s13347-017-0293-z.

Eubanks, Virginia. 2018. *Automating inequality: How high-tech tools profile, police, and punish the poor*, New York: St. Martin's Press.

European Parliament. 2017. Motion for a European Parliament resolution with recommendations to the Commission on Civil Law Rules on Robotics (2015/2103(INL)). http://www.europarl.europa.eu/doceo/document/A-8-2017-0005_EN.pdf. Accessed 22 March 2019.

Gartner. 2017. Gartner Says By 2020, Artificial Intelligence Will Create More Jobs Than It Eliminates. Press release. https://www.gartner.com/en/newsroom/press-releases/2017-12-13-gartner-says-by-2020-artificial-intelligence-will-create-more-jobs-than-it-eliminates. Accessed 22 March 2019.

Goertzel, Ben, and Cassio Pennachin, eds. 2007. Artificial General Intelligence. Berlin, Heidelberg: Springer-Verlag.

Google assistant. 2018. https://assistant.google.com/. Accessed 26 June 2018.

Google PAIR. 2018. https://ai.google/research/teams/brain/pair. Accessed 26 June 2018.

Grimmelmann, James. 2005. Regulation by software. *Yale Law Journal* 114 (7): 1719-58.

Hawking, Steven. 2014. *BBC*, December 2. https://www.youtube.com/watch?v=fFLVyWBDTfo.

Hepburn, Jarrod. 2012. The Duty to give reasons for administrative decisions in international law. *The International and Comparative Law Quarterly* 61(3): 641-663. https://doi.org/10.1017/S0020589312000309.

Horcher, Gary. 2018. Woman says her Amazon device recorded private conversation, sent it out to random contact. *KIRO7*, May 25. https://www.kiro7.com/news/local/woman-says-her-amazon-device-recorded-private-conversation-sent-it-out-to-random-contact/755507974.

House of Commons. 2018. Minutes of Proceedings, May 2. https://hansard.parliament.uk/commons/2018-05-02/debates/BE9DB48A-C9FF-401B-AC54-FF53BC5BD83E/BreastCancerScreening.

Kemp, Richard. 2016. Legal aspects of Artificial Intelligence. Kemp IT Law. http://www.kempitlaw.com/wp-content/uploads/2016/11/Legal-Aspects-of-AI-Kemp-IT-Law-v1.0-Nov-2016-2.pdf. Accessed 22 March 2019.

Lazaratos, Panagiotis. 1989. Rechtliche Auswirkungen der Verwaltungsautomation auf das Verwaltungsverfahren. Doctoral diss., University of Tübingen, Germany. https://www.didaktorika.gr/eadd/handle/10442/4608. Accessed 22 March 2018.

Medina, Eden. 2015. Rethinking algorithmic regulation. *Kybernetes* 44 (6/7): 1005-19. https://doi.org/10.1108/K-02-2015-0052.

Ng, Vivian. 2017. Algorithmic Decision-Making and Human Rights, Human Rights. Big Data and Technology Project. https://www.hrbdt.ac.uk/algorithmic-decision-making-and-human-rights/. Accessed 22 March 2018.

O'Neil, Cathy. 2016. *Weapons of math destruction: How big data increases inequality and threatens democracy*, New York: Crown Publishers.

Panetta, Kasey. 2017. Top Trends in the Gartner Hype Cycle for Emerging Technologies, 2017. Gartner. http://www.gartner.com/smarterwithgartner/top-trends-in-the-gartner-hype-cycle-for-emerging-technologies-2017/. Accessed 22 March 2019.

Pasquale, Frank. 2015. *Black Box Society*. Harvard: Harvard University Press.

Pasquale, Frank. 2016. The Emerging Law of Algorithms, Robots, and Predictive Analytics. https://balkin.blogspot.gr/2016/02/the-emerging-law-of-algorithms-robots.html. Accessed on 8 June 2018.

Peled, Roy, and Yoram Rabin. 2011. The Constitutional Right to Information. *Columbia Human Rights Law Review* 42(2): 357-401.

People + AI Research (2018). Google AI. https://ai.google/research/teams/brain/pair. Accessed on 28 June 2018.

Sartor, Giovanni, Monica Palmirani, Enrico Francesconi, and Maria Angela Biasiotti (eds.). 2011. *Legislative XML for the semantic web: principles, models, standards for document management* (vol. 4). Dordrecht: Springer.

Schwab, Klaus. 2016. *The fourth industrial revolution*. Geneva: World Economic Forum.

Selbst, Andrew D., and Julia Powles. 2017. Meaningful information and the right to explanation. *International Data Privacy Law* 7.4: 233-242. https://doi.org/10.1093/idpl/ipx022.

Shuman, Samuel I. 1971. Justification of Judicial Decisions. *California Law Review* 59: 715-732. https://doi.org/10.15779/Z38QN1W.

Stone, Peter et al. 2016. Artificial Intelligence and life in 2030. One Hundred Year Study on Artificial Intelligence. Stanford University. Report of the 2015 Study Panel. https://ai100.stanford.edu/sites/g/files/sbiybj9861/f/ai_100_report_0831fnl.pdf. Accessed 20 March 2019.

Varshney, Kush. 2018. Introducing AI Fairness 360. IBM research blog. https://www.ibm.com/blogs/research/2018/09/ai-fairness-360/. Accessed 4 June 2019.

Wachter, Sandra, Brent Mittelstadt, and Chris Russell. 2018. Counterfactual Explanations Without Opening the Black Box: Automated Decisions and the GDPR. *Harvard Journal of Law & Technology* 31 no.2 (Spring):841-887.

Wachter, Sandra, Brent Mittelstadt, and Luciano Floridi. 2017. Why a Right to Explanation of Automated Decision-Making Does Not Exist in the General Data Protection Regulation. *International Data Privacy Law* 7(2): 76-99.

Wierzynski, Casimir. 2018. The Challenges and Opportunities of Explainable AIIntel AI. https://ai.intel.com/the-challenges-and-opportunities-of-explainable-ai/. Accessed 22 March 2019.

Williams, Betsy Anne, Catherine F. Brooks, and Yotam Shmargad. 2018. How Algorithms Discriminate Based on Data They Lack: Challenges, Solutions, and Policy Implications. *Journal of Information Policy* 8: 78-115. https://doi.org/10.5325/jinfopoli.8.2018.0078.

Yannopoulos, Georgios. 2012. Technology beats the law*?* In *Values and Freedoms in Modern Information Law and Ethics*, eds. Maria Bottis, Eugenia Alexandropoulou, and Ioannis Iglezakis, 1024-1031. Athens: Nomiki Bibliothiki.

# 3 Administrative and Judicial Decisions on Advanced Algorithms

Fotios Fitsilis

**Abstract**

This chapter constitutes the main chapter of the book, where landmark cases of algorithmic regulation are presented. In total, eight cases of algorithmic conduct are discussed. These involve both direct and indirect examples of regulation. Among others, examples from the ICT sector (data-protection, internet browsers and ad-blockers) and the gig economy (i.e. Airbnb and Uber) are presented and discussed. Recent regulatory action, such as in the cases of the diesel emissions scandal and high-frequency trading are also analyzed.

Keywords: Media player, Dieselgate, Airbnb, Uber, MiFID, GDPR.

## 3.1 Microsoft Media Player and Explorer

In May 1998, the US Department of Justice (DoJ) and the Attorneys General of 20 States, as well as the District of Columbia, sued Microsoft.[1] Among others, the allegations were related to anti-competition practices through Microsoft's bundling together of its internet browser, Internet Explorer (IE) with its Windows operating systems, an action that was considered illegal under Sections 1 and 2 of the Sherman Antitrust Act.[2] The District Court of Columbia found Microsoft guilty of anticompetitive practices and ordered a series of conduct restrictions, along with an extremely invasive structural remedy of splitting Microsoft in two 'Baby Bills'.[3] The plaintiffs attempted to send Microsoft's appeal to the Supreme Court, but the court declined to render an opinion and the case was handled at a federal level.[4] A settlement was reached between Microsoft, the DoJ and nine states, which was approved by the United States Court of Appeals for the District of Columbia Circuit. The District of Columbia and nine other states rejected this settlement seeking further remedial measures, most of which were eventually rejected by the court.[5]

---

[1] US DOJ Complaint 98-12320, 1998.

[2] The Sherman Antitrust Act (1980) is included in Title 15 USC. Section 1 of the Sherman Act mentions that '[e]very contract, combination in the form of trust or otherwise, or conspiracy, in restraint of trade or commerce among the several States, or with foreign nations, is declared to be illegal'. Section 2 mentions that '(e)very person who shall monopolize, or attempt to monopolize, or combine or conspire with any other person or persons, to monopolize any part of the trade or commerce among the several States, or with foreign nations, shall be deemed guilty of a felony'.

The settlement mentions that:

'[o]riginal Equipment Manufacturers (OEMs) must be free to install and display icons, shortcuts and menu entries for competing software in any size or shape, and, along with end users, can designate the competing software as a default'.[6]

Yet, the consent decree did not require Microsoft to alter its code by removing IE. The company was only required to allow OEMs the ability to replace IE with a competing browser. The exclusion of IE from the Add/Remove Programs utility is also in violation of the Sherman Act, Section 2. Therefore the settlement provides that:

'(…) on the desktop or Start menu, OEMs and end users must be free to enable or remove access to both Microsoft's and competing software by altering icons, shortcuts, or menu entries'.[7]

Furthermore, Microsoft was prohibited from automatically altering OEMs or end user configurations through its operating system software, without confirmation from the user at least two weeks from the initial use of the computer. In addition, the OEMs ability to exercise those rights should not be restricted through licensing terms. Finally, one interesting aspect of the terms of decree is that in this case, due to the nature of rapid change in the high-technology industry, the compliance period was shortened by half (5 years instead of 10).[8]

A complaint by Sun Microsystems about the behavior of the Microsoft Corporation, with regard to its non-disclosure of protocols for Windows NT in 1998, initiated the European Commission's investigation into the company.[9] Microsoft's behavior resulted in the inability of Sun Microsystems's server products to achieve full interoperability with Windows desktop products. Hence, consumers needed to purchase both desktop and server products from Microsoft. The investi-gation was expanded in 2001 during the tying of Windows Media Player to the Window's PC operating system.[10] Due to the company's domination of the market for operating systems, this action was bound to lead in a distortion of the competition, since package-deal prices would urge consumers to favor Microsoft's products over those of competing companies.

---

[3] See Economides (2001), p. 28.

[4] *United States v. Microsoft Corp.*, 98-1232 (CKK), 231 F. Supp. 2d 144 (D.D.C.2002).

[5] *New York v. Microsoft Corp.*, 98-1233 (CKK), 209 F. Supp. 2d 132 (D.D.C. 2002); Harty and Wrobel (2003), pp. 200-201.

[6] See Cohen (2004), p. 340.

[7] See Cohen (2004), p. 341.

[8] See Cohen (2004), p. 344.

[9] See Gitter (2004), p. 187.

In its decision of March 2004,[11] the European Commission concluded that Microsoft had to disclose interface information for its server products in order for full interoperability of competitors to be achieved and enforced the creation and sale of the Windows operating system' version without Windows Media Player, named 'Windows XP N'.[12] Microsoft appealed the decision in the Court of First Instance.[13] The court's final judgement (issued on 17 September 2007) basically approved the European Commission's decision and deemed compulsory licensing as an appropriate remedy in cases of violation of Art. 82 European Communities Treaty (ECT), if certain conditions apply.[14] This also applies in the case of the technology industry, where reliance on intellectual property is fundamental to propel research and development.

Art. 81 and 82 ECT, which have been replaced by Art. 101 and 102, respectively, of the Treaty on the Functioning of the European Union (TFEU),[15] appear to be similar to Sections 1 and 2 of the Sherman Act, nevertheless different approaches have been followed to determine anti-competitive behavior. American courts use rules of reason, while under European Law behavior satisfying the conditions of Art. 81 §1 is deemed anticompetitive, unless it qualifies for an exception, as described in Art. 81 §3.[16]

During the early '90s judicial procedures against Microsoft in Europe and US managed to achieve a level of coordination.[17] This has been a step forward in international high technology antitrust cooperation. Nonetheless, Europe's more broad social welfare approach to competition law is bound to result in further action than that taken by US courts. Yet, full international cooperation in the future is essential due to the trans-jurisdictional effect of the high technology market. In a concluding note, it is suggested that developments in software technology will spare governments the need to 'interfere early in the standardization process, in the effort to ensure that the best new product wins', hence maintaining a healthy competition framework.[18]

---

[10] See Cohen (2004, p. 352.

[11] See EC Decision of 24 March 2004, Case COMP/C-3/37.392.

[12] See Cohen (2004), p. 352.

[13] *Microsoft v. Commission of the European Communities*, ECLI:EU:T:2007:289.

[14] See Gitter (2004), p. 192.

[15] Art. 101§1 of the TFEU reads as follows: 'The following shall be prohibited as incompatible with the internal market: all agreements between undertakings, decisions by associations of undertakings and concerted practices which may affect trade between Member States and which have as their object or effect the prevention, restriction or distortion of competition within the internal market, (…)'; Art. 102 TFEU: Any abuse by one or more undertakings of a dominant position within the internal market or in a substantial part of it shall be prohibited as incompatible with the internal market insofar as it may affect trade between Member States. (…).

[16] See Cohen (2004), p. 353.

[17] See Klein and Bansal (1996), pp. 179-180.

## 3.2 Volkswagen Emissions Case

The Volkswagen (VW) emissions scandal, sometimes also referred to as 'Dieselgate', is an example of an 'inverse' algorithmic regulation, i.e. a type of regulation where the VW Group used algorithms negatively and covertly, in order to affect the outcome of pollution-control calibrations and measurements. Because of its significance, level of fine and the range of implications for the car industry, the VW emissions scandal has quickly risen to a case study for academic and business purposes. It all started in 2015, when US Environmental Protection Agency (EPA) announced that it had found evidence that certain car models with diesel engines under testing conditions used an algorithmic switch, hereafter 'switch', to reduce gas emissions.[19] These switches constitute Auxiliary Emission Control Devices (AECDs) that were not previously declared, in order for the vehicles to obtain a so-called Certificate of Conformity (COC). However, when the cars operated under normal conditions, the switches were turned to a 'separate "road calibration" which reduced the effectiveness of the emission control system (…)'.[20]

According to US EPA, switches that are 'neither described nor justified in the applicable COC applications (…) are illegal defeat devices' and referred the matter to the US DoJ.[21] Both US and EU use a nearly identical language for the definition of a defeat device.[22] A second notice of violation from the EPA to the VW

---

[18] See Cohen (2004), p. 364.

[19] See relevant first 'Notice of Violation' letter to the VW Group by US EPA (2015a).

[20] See EPA (2015a), *supra* note 19, p. 4.

[21] See EPA (2015a), *supra* note 19, p. 4.

[22] A defeat device under U.S. regulation (40 CFR §86.1803-01) is 'an auxiliary emission control device (AECD) that reduces the effectiveness of the emission control system under conditions which may reasonably be expected to be encountered in normal vehicle operation and use (…)'. Similarly, under Regulation (EC) (No 715/2007 Art. 3, par. 10) 'defeat device' means any element of design which senses temperature, vehicle speed, engine speed (RPM), transmission gear, manifold vacuum or any other parameter for the purpose of activating, modulating, de-

Group followed a few months later with allegations for a defeat device in additional car models.[23] In January 2016, the US DoJ filed a civil complaint against the VW Group for alleged Clean Air Act violations.[24] Moreover, the case has triggered massive civil class action and investor lawsuits in the US and elsewhere.[25] The US DoJ also launched a criminal investigation against the VW Group on the matter. A year later, in January 2017, a preliminary framework settlement was reached in the US. VW pled guilty and agreed to pay $4.3 billion in criminal and civil penalties, for US residents only. In addition, six VW executives and employees have been charged with:

> '(…) participating in a conspiracy to defraud the United States and VW's U.S. customers and to violate the Clean Air Act by lying and misleading the EPA and U.S. customers'.[26]

In a separate settlement approved by a district Judge in San Francisco in October 2016, VW agreed to pay $14.7 billion to US consumers and government agencies.[27]

In relation to the VW emissions case, it is of importance to present and analyze the relevant reactions from the European institutions level. In late 2015, following rapid developments in the case in the US, the EP set up a 'Committee of Inquiry into Emission Measurements in the Automotive Sector'.[28] On 2 March 2017 the Committee presented its report with recommendations, which, among others, called for the forming of a new European agency to oversee road transport with market surveillance powers,[29] similar to the US EPA. However, EU member states were split over the idea of having yet another costly and probably over-bureaucratic EU agency and declined the proposal. Instead, the Members of the EP (MEPs) voted to strengthen the EC and to provide it with more powers to oversee the automotive industry.

---

laying or deactivating the operation of any part of the emission control system, that reduces the effectiveness of the emission control system under conditions which may reasonably be expected to be encountered in normal vehicle operation and use'.

[23] See second 'Notice of Violation' letter to the VW Group by US EPA (2015b).

[24] See relevant press release by US DoJ (2016); in particular, sections 204 and 205 of the Clean Air Act are mentioned therein.

[25] For a more extensive, but not exhaustive, list of lawsuits triggered by the VW diesel affair see Juul (2016).

[26] See relevant press release by US DoJ (2017) as well as the court documents therein.

[27] See VW "Clean Diesel" MDL (multidistrict litigation) on the US District Court (2018) portal of the Northern District of California.

[28] See Decision 2015/3037(RSO) of the European Parliament (2015).

[29] See the Committee report 2016/2215(INI) of the European Parliament (2017), incl. findings and conclusions.

In 2016, after talks with the EC, Volkswagen committed to repair all its cars affected by the Dieselgate scandal by Autumn 2017.[30] An additional 2-year warranty offer was given to VW owners, with the company refusing any financial compensation, arguing that a software update in its defeat devices would be enough to resolve the issues, without any loss of the car's performance.[31] This difference between VW's behavior towards its customers, i.e. cash payout in the US v. warranty extension in the EU, stems from the different legal framework in Europe, where the responsibility to approve new car models and monitor car manufacturers lay, at the time, with member states.[32]

A paradigm shift is seen in a new EU regulation, which imposes stricter requirements for emission tests and heavy fines for violations by the industry.[33] The new regulation for testing new vehicle models to be introduced on the EU market will apply as of 1 September 2020. Consequently, both national authorities and the European Commission will have the power to conduct compliance verification checks. The regulation, therefore, updates an existing directive that established a framework 'for the approval of motor vehicles and their trailers, and of systems, components and separate technical units intended for such vehicles'.[34]

## 3.3 Ad-blocking

People spend a considerable amount of their professional or free time online. Knowledge of browsing activity of data subjects is hence important to both marketing specialists and product/service providers. Companies use a broad set of tools to collect and process information about online behavior, such as computer cookies or just 'cookies'.[35] The advertising activities that are based on the analysis of personal online patterns are also known under the term 'Online Behavioral Advertising' (OBA). Several users who sensed, in the collection of their online activity data, a breach of their privacy, began using special software tools that limit OBA. The use of such tools is not a walk in the park for users, as it has been already demonstrated that such software may have serious inherent usability flaws.[36]

---

[30] See relevant press release by the European Commission (2017).

[31] See relevant Reuters report by De Carbonnel (2017).

[32] See De Carbonnel (2017), *supra* note 31.

[33] Regulation (EU) 2018/858 of 30 May 2018 on the approval and market surveillance of motor vehicles and their trailers, and of systems, components and separate technical units intended for such vehicles.

[34] This is the Framework Directive 2007/46/EC.

[35] Cookies are files stored on a user's computer. When a website is visited, these computer programs store specific activity and client-related data, which allows the server to deliver user-tailored content.

[36] See, e.g., Leon et al. (2012).

In addition, a lack of sufficient knowledge about tracking technology is further deteriorating the efficiency of existing privacy tools.[37]

OBA is a major source of financing for private online media, such as the Spiegel Online news portal,[38] contrary to public broadcasters that are financed either directly through the state budget or through legally guaranteed license fees. Income is mainly generated through advertising on stationary and mobile internet pages, within dedicated iOS/android apps, embedded videos and social networks.[39] Additional income, although on a much smaller scale, may be generated through so-called 'native advertising'.[40] This is a new means of web-advertising that is thought to offer added-value to the users of online content. Native ads are designed to look and feel similar to the surrounding media format and suggest related and clearly indicated content to website visitors. 'Affiliate links' are also indicated as such in host websites.[41] These are Uniform Resource Locators (URLs) that contain the affiliate's ID or username. For instance, in the case of a book review, a space that contains details to a particular book, i.e. Publisher, page number and price, may be displayed as well as links to the online retailers from which the book can be purchased. Should someone click on those ads and make a purchase, the provider of the original website gets a commission.

The blocking of OBA through ad blockers and anti-tracking software has sparked several judicial disputes, since it can be used to blend out several types of the aforementioned online ads. Consequently, the overall advertising revenues sink.[42] One of the most significant cases has been recently decided in Germany.[43] The German company Eye/o GmbH, hereafter Eyeo, offers online users a free version of the ad blocker AdBlock Plus. This software filters ads that are listed in a so-called 'Black List'. Eyeo offers companies a possibility to circumvent this blockade by entering their ads into a 'White List'. In order for the companies to do so, their ads need to fulfill a certain set of criteria and they need to share online revenues with Eyeo. The company says it excludes Small and medium-sized En-terprises (SMEs) from the revenue participation condition.[44] It is fair to assume that online media saw in this computer program a disruption to their business model and therefore proceeded with legal action against the software development company.

---

[37] Leon et al. (2012), p. 598.

[38] Spiegel Online GmbH & Co. KG, a private German media company, operates the portal; its parent company is SPIEGEL-Verlag Rudolf Augstein GmbH & Co. KG; the media company is used here solely as an OBA study case as the majority of online media uses similar advertising schemes to generate online revenue.

[39] For additional information and structured analyses on the financing of online media see, e.g., Ulin (2013) and Cea-Esteruelas (2013).

[40] More information on the effects of positioning and language in native advertising, sometimes also mentioned as 'sponsored content', may be found in Wojdynski and Evans (2015).

[41] See, e.g., Edelman and Brandi (2015).

[42] Dörting et al. (2017) offer an indication of the share of the online revenues compared to the total revenues for the case of Spiegel Online, i.e. in 2016 the share has exceeded 80%.

[43] *Axel Springer AG v. Eyeo*, I ZR 154/16, BGH.

After going through lower court instances, the case has been decided at the German *Bundesgerichtshof* (BGH, Federal Court) in Karlsruhe, which declared the internet operation of AdBlock Plus lawful. According to the original claim by the media company, Axel Springer AG, the software company violates German competition law (Gesetz gegen den unlauteren Wettbewerb, UWG)[45] and accused Eyeo of unfair competition and aggressive business practices, arguing that it threatens to destroy the online business model of advertising-financed media.[46] However, the court decided that the mentioned computer program which automatically suppresses internet ads on web pages does not violate competition law. In particular, it decided that there is no legal basis to ban software that –like the popular AdBlock Plus– allowed users to read desired content online, while opting out of digital advertising. The BGH ruling mentions that any financial damage caused was not inflicted by the software provider but by users who downloaded its software. The court also pointed out that media companies have always the possibility to exclude users that utilize ad blockers from their native content. The ruling concludes that:

> '[t]here is also no general market obstruction because there is insufficient evidence that the business model of providing free content on the internet is being disrupted'.[47]

Another related landmark decision regarding filtering of content, at the EU level this time, was reached back in 2011.[48] Here, the Court of Justice of the European Union (CJEU) found that an injunction against a Belgian Internet Service Provider (ISP), Scarlet Extended SA, to install a general filter for all electronic communications passing through its services in order to prevent illegal downloads, was against EU law. The case was ruled on a completely deferent legal basis compared with *Axel Springer AG v. Eyeo*, which was ruled on the grounds of German competition law.[49] According to the Court, the injunction resulted in a:

---

[44] See relevant press announcement by BGH (2018) on the *Axel Springer AG v. Eyeo* case, *supra* note 43.

[45] In particular, the relevant provisions for the BGH ruling are §4 Nr. 4 UWG and §4a UWG.

[46] See BGH (2018), *supra* note 43.

[47] Translation from German from BGH (2018), *supra* note 43.

[48] *Scarlet Extended SA v. SABAM*, ECLI:EU:C:2011:771.

[49] The legal basis for the *Scarlet Extended SA v. SABAM* ruling included: Directive 2001/29/EC on copyright, Directive 2004/48/EC on intellectual property rights, Directive 2000/31/EC on electronic commerce, Directive 95/46/EC on the processing of personal data and Directive 2002/58/EC on privacy and electronic communications.

'(…) serious infringement of the freedom of the ISP concerned to conduct its business, since it would require it to install a complicated, costly, permanent computer system entirely at its own expense (…)'.[50]

The Court also considered the cost dimension of the original injunction under the light of the provisions of Directive 2004/48/EC, which requires that '(…) measures to ensure the respect of intellectual property rights should not be unnecessarily complicated or costly'.[51]

## 3.4 Block Controversial Content

Google and other search engines have acquired immense power as public opinion shapers, a fact which eventually goes hand in hand with the responsibility for presented and promoted content. Modern search engines embed sophisticated algorithms to improve and speed up user searches, such as the algorithms for the autocomplete and suggest mechanisms. Based on their implementation, such algorithms foresee the next word(s) a user is likely to type according to local, regional or global search criteria. In this respect, it has been argued that issues of intellectual property or personal data cannot be left to be regulated by a random internet majority at a certain moment.[52]

On 13 May 2014 the CJEU ruled on a dispute between Google[53] and the Spanish data protection agency 'Agencia Española de Protección de Datos' (AEPD), as well as Mr. Mario Costeja González, a Spanish national resident in Spain, on the right of the latter to ask Google to remove results for queries that include the person's name.[54] In this landmark case the story unfolds as follows. In 2010, Mr. González and the AEPD filed a complaint against a daily newspaper in Catalonia (Spain), which is owned by La Vanguardia Ediciones SL, hereafter 'La Vanguardia', Google Spain and Google Inc.

The reason for the complaint was based on the fact that the Google search engine, when prompted with Mr. González's name, returned links on two specific pages of La Vanguardia's newspaper. These pages dated back to 1998 and mentioned Mr. González's name in relation to a real-estate auction along with his connection to attachment proceedings for the recovery of social security debts prior to that time. Mr. González requested from La Vanguardia either to remove or alter those pages in order for his personal data to no longer be visible or, alternatively, to use appropriate tools offered by search engines to protect his personal data. Additionally, he turned to the Google search engine requesting by Google Spain or Google Inc. to either remove or conceal personal data relating to him, so as to stop appearing in related search results such as the links to La Vanguardia newspaper. In this respect, he argued that the mentioned attachment proceedings had been fully resolved years ago, making any reference to them completely irrelevant.

---

[50] *Scarlet Extended SA v. SABAM*, *supra* note 49.

[51] *Scarlet Extended SA v. SABAM*, *supra* note 49.

[52] See Yannopoulos (2013).

[53] Google is registered trademark of Google LLC; its parent company is Alphabet Inc.

[54] *Google Spain SL and Google Inc. v. Agencia Española de Protección de Datos (AEPD) and Mario Costeja González*, ECLI:EU:C:2014:317.

The case has been first ruled before the Audiencia Nacional, the Spanish National High Court. The Court stated that the legal action raised the question of which obligations search engine operators have, in order to protect personal data of persons who do not wish certain personal information on third parties' websites:

- to be linked back to them,
- to be located,
- to be indexed and
- to be made available to internet users indefinitely.

The court recognized that answering the above question is linked to the relevant interpretation of Directive 95/46/EC.[55] The latter needs to address the underlying technologies which have been developed after its publication in 1995. In the light of the above, the court decided to refer to the CJEU for a preliminary ruling. The CJEU used Directive 95/46/EC as well as the Charter of Fundamental Rights of the European Union[56] as the legal basis for its ruling, the most significant points of which are presented and analyzed in the following paragraphs.

This is a general ruling that applies beyond Google to all search engines that operate a branch or subsidiary in an EU member state. One of the most important outcomes is the assessment that search engine activity[57] constitutes the 'processing of personal data' when information that is being handled contains personal data (Art. 2(b) Directive 95/46/EC ). As a direct consequence, the operator of the search engine must be regarded as a 'controller' that processes personal data in the sense of Art. 2(d) Directive 95/46/EC. Hence, if the relevant provisions are satisfied, the CJEU ruled that the operator of a search engine is obliged to remove results that contain links to web pages that contain personal information following a search made 'on the basis of a person's name'. Interestingly, the operator must remove the results of the search engine algorithm:

> '(…) also in a case where that name or information is not erased beforehand or simultaneously from those web pages, and even, as the case may be, when its publication in itself on those pages is lawful'.[58]

---

[55] Since 24 May 2018, this directive is no longer in force. It is repealed by Regulation (EU) 2016/679 (GDPR); see *infra* Section 3.7; Art. 17 GDPR outlines the 'different circumstances under which individuals can exercise the right to be forgotten'.

[56] The Charter of Fundamental Rights of the European Union (2000/C 364/01) was published in 18 December 2000 in the Official Journal of the European Communities.

[57] *Google Spain SL and Google Inc. v. AEPD and González*, *supra* note 54; the ruling mentions that search engine activity includes the finding, automatically indexing, temporally storing and the making available to internet users of information on the internet by third parties.

What arises as an issue of widespread importance from the aforementioned ruling is the fact that the exercise by a data subject of 'the right to be forgotten'[59], meaning that its relevant personal information will no longer appear in a publicly available search engine list, overrides, as a general rule both the 'economic interest of the operator of the search engine' and the 'interest of the general public' in having access to that information'.[60] However, a balance is to be held when exercising these rights.[61] For instance, compared to the average citizen, the 'right to be forgotten' of a data subject with a public function or a significant role in public life is to be assessed in a different manner, stand-alone or vis-à-vis the other mentioned rights. Nevertheless, who is to decide in which cases the 'right to be forgotten' is to be granted? Certainly, in the first instance, the search engines are the ones which absorb the entire workload of assessing numerous complaints from the data subjects. To decide which request qualifies, search engines must examine the related results for being either 'inadequate, irrelevant, no longer relevant, or excessive'.[62] In the case of a disagreement between the data subject and the search engine, the local Data Protection Agency (DPA) is to be consulted before further legal action is considered.[63] Finally, the courts, national or supra-national, constitute the last resort where any disputes shall settled.

---

[58] *Google Spain SL and Google Inc. v. AEPD and González*, *supra* note 54.

[59] Sometimes also referred to as the 'right to erasure'.

[60] Here is reference to the 'right to operate freely', meaning that the operator of the search engine should be able to operate the company in a financially viable manner, and to the 'right to information' of the general public in having access to a list of results that a search engine algorithm returns.

[61] The CJEU has been criticized for its failure to achieve a 'fair balance' between conflicting rights; see, e.g., Frantziou (2014), p. 768; The European Court of Human Rights has also dealt with conflicts between the right to privacy and the right of freedom of expression; overall, a high level of coherence in the relevant case law between the two major European courts is detected; see Margaritis (2018); The case, *M.L. and W.W. v. Germany*, ECLI:CE:ECHR:2018:0628JUD006079810, further elucidates the state of play on privacy issues in Europe; the judgment provides strong protection of media archives, while being consistent with the CJEU reasoning for search engine operators in the *Google Spain SL and Google Inc. v. AEPD and González* case.

[62] For instance, Google (2019) shows how a major search engine operator implements 'the right to be forgotten' and the way it copes with individual removal requests.

The new Regulation (EU) 2016/679 (GDPR) incorporates the essence of the aforementioned CJEU ruling in Art. 17, titled 'Right to Erasure ('right to be forgotten')'. The root of this right can be traced back to French law, which recognizes 'le droit à l'oubli', which allows rehabilitated criminals to object to the publication of the facts of their conviction and incarceration.[64] Under GDPR, in exercising a data subject's 'right to be forgotten', personal data needs to be erased from the controller's environment, taking into account 'available technology and the cost of implementation'. Cost considerations within regulatory provisions are now new. They have been also part of previous regulatory provisions and have been invoked in court rulings.[65]

Also, the controller would have to make sure to erase any publicly made links to information containing personal data. Additionally, data subjects can request erasure of their personal information when a) this is no longer required for the purpose it was collected, b) there is no legal basis for processing it and c) consent has been withdrawn. However, as seen in the above CJEU ruling, the GDPR also mentions some significant exceptions to the 'right to be forgotten'. For instance, in cases where processing is necessary:

- to exercise the right of freedom of expression and information,
- to comply with an EU or a member state legal obligation,
- to perform a public interest task or exercise of official authority,
- for public health reasons or
- for archival, research or statistical purposes.

Implementation of the 'right to be forgotten' has serious implications for data controllers and processors, not only to search engines, but to the entity of organizations[66] which at any point of their operation involve collection and processing of personal data, by own means or through third parties. Recent guidelines on profiling and individual decision-making point out that the right to erasure applies to both the input and output data, i.e. the personal data to create a profile and the re-sult of the profiling process, respectively.[67] Consequently, organizations need to have in place appropriate legal and technical procedures to be able at any time to erase, upon request, any of the personal data held. As previously seen, compliance to this right requires a case by case assessment from the side of an organization, in order to ensure that the handling of personal information is considered appropriately.[68] As this right in not absolute, data controllers also need to include in their decisions, several, sometimes conflicting, factors such as technology, costs and proportionality.

---

[63] See Google (2019), *supra* note 62.

[64] See Rosen (2011), p.88.

[65] See, e.g., Directive 2004/48/EC, which requires that measures to protect intellectual property rights are - among others - not to be costly, and the Scarlet Extended SA vs. SABAM case, which is discussed in *supra* Section 3.3; further cost considerations when regulating the internet are presented by Goldsmith and Sykes (2001) under the light of the US Dormant Commerce Clause.

[66] The term 'organization' covers here any form a legal entity may take, e.g. personal company, SME, private/public company, agency, NGO, multi-national corporation etc.

## 3.5 Sharing economy

The use of software platforms in the sharing economy is directly linked with efficiencies deriving from reducing transaction costs, the encouraging of accountability and competition, improving allocation of resources, and information and pricing advantages.[69] In the following paragraphs, a number of significant cases around two of the main actors of the sharing economy, Airbnb and Uber, are discussed.

### 3.5.1 The Airbnb case

Airbnb is a popular online rental platform, which has evolved into a global phenomenon during its ten years of operation (2008 to date). Today, the website lists nearly five million residences in over 191 countries and 81,000 cities worldwide, while more than 500 million visitors have used its services.[70] As a result of its rapid expansion and its inherent dynamic, the service has stirred up local markets, particularly the real estate and hotel industries and a multitude of cities and countries have been urged to impose restrictions on the platform's operation, e.g. in terms of the maximum number of properties per owner or the maximum number of days per year that a property can be made available through the platform. Typical examples are most European metropoles, such as Paris, Amsterdam, Berlin, Barcelona, as well as regions that attract a high number of foreign visitors during the summer period. However, the need to regulate the market for short-term leases

---

[67] See WP29 (2018); WP29 is an advisory body (working party) on data protection and privacy set up under Art. 29 of Directive 95/46/EC.

[68] For this an organization needs to be able to assess the type of information, its sensitivity for the data subject and the interest of the general public.

[69] See, e.g., Edelman and Geradin (2016); in this relation, see also Interian (2016), pp. 130 and 137.

[70] See Airbnb (2019a).

is more pronounced in urban areas that are already facing housing problems or shortages of supply.[71]

The activity of sharing platforms has caused several judicial disputes across the globe. A typical example involves the City of Paris taking multiple legal action, primarily against Airbnb, but also against other providers of similar services, such as Wimdu. With reference to housing shortages, the city administration in Paris has long tried to enforce strict regulatory measures upon online platforms. In the above court case and via express procedure, the municipality of Paris asked the court to fine the platforms for not removing ads lacking a proper registration number,[72] which is mandatory to every rented property in Paris.[73] Additionally, the city has imposed a maximum of 120 days per year for rented spaces as accommodation for tourists.[74] Moreover, the *Assemblée Nationale* back in 2016 decided that mediation portals must report on a yearly basis the number of transactions and the revenues of their users directly to the national treasury department. The measure is to enter into force in 2020 for income generated in 2019.[75]

In a prominent case, the City of Paris asked the court to fine the platforms for not removing ads lacking a proper registration number, which is mandatory to every rented property.[76] The result of the hearing before the *Tribunal de Grande Instance (TGI) de Paris* (High Court of Paris) was expected with high anticipation to take place on 12 June 2018. However, Airbnb posed a question on a 'priority issue of constitutionality'[77] claiming violation of the constitutional principle of equality before public office, which was considered positively by the President of the court, who referred the subject to the Court of Cassation. However, the court rejected the claim, stating that it does not have to refer the matter to the Constitu-tional Council.[78] In another similar case, in front of the *TGI de Paris*, the City of Paris accused Airbnb of displaying properties without registration number. In its decision, the court dismissed the claim considering that the absence of a registration number for limited period rentals does not constitute unlawful behavior.[79]

---

[71] In Greece, *Nomos* (Law) 4472/2017 (Art. 84) fully liberalized the activity of sharing platforms, such as Airbnb. However, restrictions may be imposed locally by ministerial decisions, e.g. in the case of housing problems.

[72] See relevant Reuters report by Rivet (2018).

[73] See relevant French 'Law for a Digital Republic' (*Loi n° 2016-1321 du 7 Octobre 2016 pour une République numérique*) and in particular Art. L.324-1-1 of the Tourism Code introduced therein.

[74] See relevant decision (2017 DLH 128) by the Mayor of Paris.

[75] See amended 2016 French budget law (*Loi n° 2016-1918 du 29 Décembre 2016 de finances rectificative pour 2016*) together with the French 'Anti-Fraud Act' (*Loi n° 2018-898 du 23 Octobre 2018 relative à la lutte contre la fraude*); in this regard it is interesting to note that Art. 24 of the amended French budget law for 2016 targets *opérateurs des plateformes en ligne* (online-platform operators) in general, according to Art. L. 111-7 *du code de la consummation* (consumer code), thus allowing for a broad interpretation across business sectors and regardless of the seat of a legal entity.

[76] See *supra* note 73.

[77] English for *Question Prioritaire de Constitutionnalité (QPC)*.

Apart from its dispute with the digital platforms, the City of Paris has also engaged directly multi-owners who do not declare their properties. Facing millions in fines, owners appealed in cassation before the Court of Cassation, arguing that the relevant sanctions by French law do not comply with European law.[80] In the meantime, several legal proceedings initiated by the City of Paris before Paris courts are suspended awaiting the CJEU ruling,[81] which is expected to be pronounced in late 2019 or at the beginning of 2020.

This kind of lengthy multi-instance and multi-dimensional legal battles around the operations of Airbnb and its peers are also taking place elsewhere in Europe. Gradually, it becomes clear that metropolitan cities, rather than the state (or the European Union), are the most active peers in imposing regulation on the activities of the digital platforms. The reasons for that are at hand, such as a deteriorating housing situation that can be monitored more efficiently at the local level, as local self-government is more likely to feel directly the effects of the gig economy, be it positive or negative.

Just like in the occasion of the City of Paris, several other major European cities are considering imposing a similar regulatory framework against sharing platforms such as Airbnb. As a basic measure, authorities limit the days per year a location/accomodation can serve as transient accommodation. In Amsterdam, the city council wants to tighten the rules for renting apartments to tourists via online platforms. Starting in 2019, landlords will be allowed to rent their apartments only for 30 days per year, half of the previous maximum of 60 days per year. Similar to Paris, there is also an obligation for landlords to notify their municipality in advance should they offer their home for renting through an online housing platform.[82]

In comparison to their US counterparts, European cities have presented faster reflexes to remedying home sharing externalities.[83] A similar measure has also been adopted in Japan. The 'Minpaku Shinpou', the new law for the private lodg-ing business, took effect as of 15 June 2018 and is considered to be the first relevant national legal framework in Asia.[84] Minpaku limits home-sharing to 180 days a year and requires homeowners to notify the city government and to be subject to additional control measures. As a consequence of minpaku, Airbnb informed homeowners that their properties would not be listed on the platform, unless they are in compliance with the new law. The fact that platform software needs to be adapted to check whether a landlord or a property is appropriately registered constitutes a clear and high-profile case of algorithmic regulation.

---

[78] *Airbnb France v. City of Paris*, ECLI:FR:CCASS:2019:C300154.

[79] *City of Paris v. Airbnb France and Airbnb Ireland,* TGI Paris, n° 18-54.632.

[80] *City of Paris v. Ms. Claire G.F.*, ECLI:FR:CCASS:2018:C301005.

[81] The preliminary question to the CJEU ruling is related to the conformity of Article L.631-7 of the French Construction Code with the provisions of Directive 2006/123/EC on services in the internal market.

[82] See Airbnb (2019b) for an overview of the home sharing issues in the City of Amsterdam; Dredge et al. (2016) present a comprehensive analysis of regulatory approaches for four major European Cities, i.e. Barcelona, Berlin, Amsterdam and Paris.

[83] See Interian (2016), p. 157.

In view of the worldwide regulatory activity around sharing platforms, an interesting response from the companies may be observed that could be part of a future generalized trend. At least in a couple of occasions, the private companies that operate the digital platforms have opted to positively respond to authorities' demands.[85] On the motives of such rather novel corporate behavior may be debated. Presumably, being pro-active will not help them evade further legal or administrative action but consumers, administrators or the judiciary could indeed regard it as a move of goodwill. This again could translate in buying time and saving large amounts in fines.

The fact that Airbnb operates a digital services platform does not mean that any regulations targeting it constitute *per se* cases of 'direct regulation'. Take for instance the decrease in the number of days a property can be made available to home sharing. In this case, regulators mainly attempt to tackle an underlying issue, e.g. provide solutions the housing issue in metropolitan areas or protect a traditional economy sector such as the hospitality industry (or both). This type of regulation is named 'indirect regulation'. We speak of the alternative option, i.e. 'direct regulation', when modifying the actual digital platform. Such modifications may induce changes in the periphery software, the user interface or the primary algorithm.[86] Hence, the case where platform software needs to be adapted to check whether a landlord or a property is appropriately registered constitutes a case of direct algorithmic regulation.

---

[84] See, e.g., Matsui (2019), pp. 130-132.

[85] Airbnb has committed to adequately respond to a respective call from the European Commission and EU consumer authorities; see, e.g., EC (2018); another pro-active move by the technology community is to be found in joining EU's Code of Practice on Online Disinformation; see, e.g., Marsden and Meyer (2019), p. 4.

[86] It is not always possible to determine whether a change in functionality is attributed to a minor software update or a major version revision that includes changes in the core algorithm. This would require a study the source code, which in the mentioned cases constitutes proprietary intellectual property and may be protected by patents, source code copyright or else; see, e.g., Liberman (1995).

### 3.5.2 The Uber case

Another significant case study of the sharing economy is to be found in the name of Uber Technologies Inc., hereafter Uber. Uber is an online platform that offers, among others, peer-to-peer personal transport services and has quickly evolved into a technology firm that deals with self-driving technology and urban air transport. Similar to the Airbnb case, there is a wide range of lawsuits worldwide concerning Uber,[87] some of them aiming to even clarify the nature of the service. In France, in particular, the law requires that each driver needs to return to the base in between rides, thus preventing the circulation of vehicles in areas with anticipated customer demand. This raises the cost of each ride and promotes traffic congestion, as more distance needs to be covered.[88]

In a landmark case, a Barcelona Court 'Juzgado de lo Mercantil No 3' requested by the CJEU a preliminary ruling regarding the nature of the Uber ridesharing service.[89] In particular, the Spanish Court sought guidance in the case *Asociación Profesional élite Taxi v. Uber Systems Spain SL* on whether the Uber service can be regarded as:

- a transport service,
- an electronic intermediary service or
- an information society service.

The CJEU judged that Uber 'intermediation services' that are delivered 'by means of a smartphone application' are inherently linked to a transport service and must be classified as 'a service in the field of transport'.[90] Hence, Uber's services

---

[87] See, e.g., Edelman and Geradin (2016).

[88] See *Loi* (Law) n° 2014-1104 du 1er Octobre 2014 relative aux taxis et aux voitures de transport avec chauffeur [Relating to taxis and chauffeur-driven transport], JORF n°0228 15938 (2 October 2014).

[89] *Asociación Profesional Elite Taxi v. Uber Systems Spain, SL,* ECLI:EU:C:2017:981.

[90] Within the meaning of Art. 58(1) TFEU. The operative part of the judgment reads as follows: 'Article 56 TFEU, read together with Article 58(1) TFEU, as well as Article 2(2)(d) of Directive 2006/123/EC of the European Parliament and of the Council of 12 December 2006 on services in the internal market, and Article 1(2) of Directive 98/34/EC of the European Parliament and of the Council of 22 June 1998 laying down a procedure for the provision of information in the field of technical standards and regulations and of rules on Information Society services, as amended by Directive 98/48/EC of the European Parliament and of Council of 20 July 1998, to which Article 2(a) of Directive 2000/31/EC of the European Parliament and of the Council of 8 June 2000 on certain legal aspects of information society services, in particular electronic commerce, in the Internal Market ('Directive on electronic commerce') refers, must be interpreted as meaning

fall outside the scope of the Directive 2006/123/EC, which implements Art. 56 TFEU on the free movement of services. Instead, Art. 58(1) TFEU needs to be applied, which relates specifically to the freedom to provide services in the field of transport. The Court recognized that there are no European common rules based on non-public urban transport services, meaning that it is up to the member states to regulate the conditions under which intermediation services are to be offered, thus creating additional national regulatory overhead with which companies such as Uber need to comply.[91]

One of the many lawsuits against Uber in the US, *Philadelphia Taxi Association Inc. v. Uber Technologies*, was ruled in the US Court of Appeals. In particular, 80 Philadelphia taxi companies that constitute the Philadelphia Taxi Association asked the Court of Appeals to reverse the order of a lower court contending that:

> '(…) Uber violated the antitrust laws because its entry into the Philadelphia taxicab market was illegal, predatory, and led to a sharp drop in the value of taxicab medallions as well as a loss of profits'.[92]

The appellants argued that traditional taxis are losing money because of unfair competition. However, the Court stated that the purpose of antitrust laws is to 'protect competition, not competitors'. Ultimately, the court decision –like other similar decisions in the US– concluded that taxi companies:

> '(…) have no right to exclude competitors from the taxicab market, even if those new entrants failed to obtain medallions or certificates of public convenience.'[93]

In a further US lawsuit filed in 2017 in front of the Central District Court in California, Uber is charged to have implemented an 'active, extensive, methodical scheme (…) to defraud drivers'.[94] Uber is pricing its services using an algorithmic

---

that an intermediation service such as that at issue in the main proceedings, the purpose of which is to connect, by means of a smartphone application and for remuneration, non-professional drivers using their own vehicle with persons who wish to make urban journeys, must be regarded as being inherently linked to a transport service and, accordingly, must be classified as 'a service in the field of transport' within the meaning of Article 58(1) TFEU. Consequently, such a service must be excluded from the scope of Article 56 TFEU, Directive 2006/123 and Directive 2000/31'.

[91] In Greece, the mentioned decision has been transposed in the national legal order via Art. 13 of Law 4530/2018 'Regulation of transport issues and other provisions' (translated from Greek).

[92] *Philadelphia Taxi Association, Inc. v. Uber Technologies*, 17-1871, US Court of Appeals.

[93] *Philadelphia Taxi Association, Inc. v. Uber Technologies*, *supra* note 92.

[94] See class action complaint, *Sorvano Van v. Rasier, LLC., Rasier-CA, LLC., and Uber Technologies, INC.*, hereafter '*Sorvano v. Uber*', case 2:17-cv-02550-DMG-JEM, US District Court for the Central District of the State of California.

pricing system, which calculates the total fare before, instead of after, a ride. According to the claim, Uber uses '(…) a base fare plus a per-mile and per-minute charge for the estimated time and distance of the travel, respectively'.[95] An Uber driver, Mr. Sophono Van, claimed that the company manipulated the software in the Uber app to calculate longer and slower routes than the typical one needed to reach a passenger's destination. Hence, according to the plaintiff of this class suit,[96] the fare for the passenger would be higher, while the driver's commission is calculated based on the typical (cheaper) route, with Uber keeping the difference. Should these allegations prove correct, then this will constitute another case of algorithmic misconduct, similar to *US v. VW*, where a defeat device was used to trick emission tests in diesel engines. In the following paragraphs, a series of general issues regarding the operation of software platforms in the sharing economy are discussed.

Concerns regarding algorithmic bias have been presented in Sections 2.1.1 and 2.2. In the above case, decentralized decisions by each host, on whether or not to accept customers, may allow for instances of discrimination. The findings of a relevant US study demonstrate that participation, pricing and ratings on Airbnb are in accordance with existing racial inequalities.[97] This happens in spite of the overall ameliorative effect that the use of the platform can have on the economy of the affected groups.[98]

The issue of platform operators' *liability* for the activities their platforms facilitate or coordinate, is still under debate. This is not a specific focus however, of this book. Nevertheless, liability represents one of the critical issues that employment will be confronted with in the sharing economy era.[99] In the US, because of the determination to create a friendlier environment for innovation, decreased liability for Internet Service Providers (ISPs) was ensured through legislation such as §230 of the Communications Decency Act (CDA). In 2018, a first relevant federal ruling on the sharing economy stated that Uber limousine drivers are independent contractors and not the company's employees within the meaning of the Fair Labor Standards Act.[100] The EU is looking forward to reform the current rules governing ISPs copyright liability beyond the Directive on electronic commerce (2000/31/EC) and the copyright Directive (2001/29/EC). After long discussions, the new Directive 2019/790 is about to enter into force.[101] The underlying discus-sion on the necessity for pan-European liability rules, as well as relevant CJEU case law, is presented in a paper ordered by the European Parliament's Committee on the Internal Market and Consumer Protection.[102]

---

[95] See points 63 and 64 in *Sorvano v. Uber, supra* note 94.

[96] This is a type of lawsuit where one of the parties is a group of people who are represented collectively by a member of that group.

[97] See Cansoy and Schor (2017), pp. 12-15.

[98] See Cansoy and Schor (2017), p. 17.

[99] See Cunningham-Parmeter (2016).

[100] *Razak v. Uber Technologies Inc.*, 2:16-cv-00573, US District Court for the Eastern District of Pennsylvania.

[101] See Directive (EU) 2019/790 on copyright and related rights in the Digital Single Market.

In addition, in the sharing economy, the issue of 'offer of universal service' is at stake. For instance, Taxi fleets and conventional hotels are required to provide a percentage of vehicles and rooms, respectively, that can accommodate wheelchairs. To a great extent, this is not possible for the owner of a single house or vehicle. The offer of incentive payments –funded through special taxes– to providers of universal service, is a proposal worth investigating.[103] Overall, it is the inherent externalities occurring by the platforms' operation that require regulatory intervention. Safety is another field of intervention. The lack of contractual relationship removes the motive for Uber drivers to obtain a commercial endorsement for their driver's license and Airbnb residences might lack proper fire escaping planning. Such issues constitute a field that should be regulated on a bare minimum requirements basis.[104]

Concluding, in the sharing economy, algorithms and innovation find plenty of room for expansion. Policymakers should embrace efficiencies generated through the operation of software platforms by removing unnecessary or protectionist rules, while, at the same time, imposing a regulatory framework to ensure the legality and safety of these platforms, adapt market operation and uphold the general public interest.[105]

## 3.6 Algorithmic trading regulation

The rise of algorithmic trading offers an interesting and developing field of algorithmic regulation. The European Union has long been at the forefront of financial market regulation. Applicable since 2007, the Markets in Financial Instruments Directive (2004/39/EC), or 'MiFID' has been the cornerstone of the EU's regulation of financial markets, but recent market anomalies[106] have made an update of the relevant framework necessary. The new legislation, MiFID II (2014/65/EU), entered into force in 2018 and is accompanied by a set of new regulations, as well as by detailed technical rules (standards) for implementation.[107] Significant im-provements include new reporting requirements and tests that will increase the amount of information available reducing the use of dark pools[108] and Over-The-Counter (OTC) trading.[109] In addition, there are stricter rules governing high-frequency-trading. In relation to algorithms, the new regulatory scheme introduces strict oversight and monitoring of algorithmic trading, by imposing new and detailed requirements on both algorithmic traders and trading venues (Art. 17 MiFID II).

---

[102] See Nordemann (2018).

[103] See Edelman and Geradin (2016), p. 321.

[104] See Edelman and Geradin (2016), pp. 309-316.

[105] See Edelman and Geradin (2016), p. 296.

[106] For instance, the 'Flash Crash" in May 2010 and the 'Tweet Flash Crash' in April 2013 are understood here as market anomalies.

[107] The word is here about the Markets in Financial Instruments Regulation (MiFIR), Market Abuse Regulation (MAR) and ESMA's Regulatory Technical Standards.

Algorithmic trading utilizes advanced algorithms to automatically determine several order parameters[110] and calculate optimal trading strategies with limited or no human intervention.[111] Although the industry draws positive feedback from the introduction of high frequency and algorithmic trading, regulators are concerned that such technologies may cause market distortion. Specific concerns involve:

> '(…) high order cancellation rate, increased risk of overloading systems, increased volatility, the ability of algorithmic traders to withdraw liquidity at any time and insufficient supervision by competent authorities'.[112]

Consequently, a company[113] that is dealing with algorithmic trading will be required to:

> '(…) have in place effective systems and risk controls (…) to ensure that its trading systems are resilient and have sufficient capacity, are subject to appropriate trading thresholds and limits and prevent the sending of erroneous orders or the systems (…)'.[114]

At the same time, trading venues need to be able to identify which orders have been placed by algorithmic trading. Both companies and trading venues need to take 'robust measures' in order to ensure that 'algorithmic trading or high-frequency algorithmic trading techniques do not create a disorderly market and cannot be used for abusive purposes'.[115] For instance, trading venues need to provide their members with algorithm testing facilities. This is an important organizational requirement, which allows companies to test their algorithms within a controlled non-live environment, similar to the one that is used for the online real-time trading (ex-ante identification of negative algorithmic impact). The relevant MiFID II provisions also lead to the implementation of systems and procedures that are able to identify any negative impact ex-post, e.g. through implementation of a 'kill button' in order to cancel any outstanding orders.

---

[108] Dark pool (also black pool) is a private forum for trading securities, derivatives, and other financial instruments.

[109] The phrase refers to stocks that trade via a dealer network rather than on a centralized exchange.

[110] Order parameters may include the time of submission, price, order management, timing of withdrawal etc.

[111] A definition of 'algorithmic trading' is also presented in MiFID II, recital 39.

[112] See Norton Rose Fulbright (2014).

[113] MiFID uses the term 'investment firm' instead.

[114] See Art. 17(1) MiFID II.

[115] See Recital 64, MiFID II.

Apart from kill buttons or switches, in order to avoid market distortion, regulators insisted in companies having specific minimum pre-trade risk limits on order submission,[116] such as:

1. price collars,
2. maximum order value,
3. maximum order volume,
4. maximum long-short positions,
5. Maximum long/short overall strategy positions,
6. repeated automated execution throttles,
7. outbound message rates,
8. maximum message limits and, where appropriate,
9. market maker protections.

For the same reason, the European Securities and Markets authority (ESMA) also proposes that companies should be able to automatically block or cancel non-eligible trade orders above the company's risk management threshold.[117]

High-Frequency Algorithmic Trading (HFAT)[118] is a subset of algorithmic trading with high message intraday rates regarding orders, quotes or cancelations.[119] Companies that utilize HFAT are required to store in an approved form 'accurate and time sequenced records of all (…) placed orders' for at least five years and '(…) shall make them available to the competent Member State authority upon request'.[120] The records should be detailed enough to be able to identify:

> '(…) the person in charge of each algorithm, a description of the nature of each decision or execution algorithm and the key compliance and risk controls'.[121]

The most significant changes to the relevant US legislation resulted from the 2010 Dodd–Frank Wall Street Reform and Consumer Protection Act.[122] Due to inherent complexity and the rather new legislation on the topic, judicial decisions with regard to algorithmic trading or HFAT are not frequent.[123]

---

[116] See ESMA (2014a), pp. 223-224.

[117] See ESMA (2014a), p. 224.

[118] For a definition of HFAT see Recital 40, MiFID II.

[119] Such a volume of intra-day messages would be in the range of 75,000 messages per trading day on average over the year, according to the ESMA (2014b), p. 231.

[120] See Art. 17(2) MiFID II.

[121] See Norton Rose Fulbright (2014).

[122] The Dodd–Frank Wall Street Reform and Consumer Protection Act became Public Law No: 111-203 in 21 July 2010.

[123] See US DoJ docket for commodities fraud.

In 2015, a US District Court, in the Northern District of Illinois,[124] convicted Mr. Michael Coscia, a high-frequency trader who was accused of commodities fraud and 'spoofing'.[125] Mr. Coscia was accused of placing in the year 2011 large orders into futures markets that he never intended to execute, thus creating an illusion of high demand, with the intent to lure other traders to markets. Spoofing was declared illegal in the commodities futures market in the US through the 2010 Commodity Exchange Act. Mr. Coscia appealed the decision in front of the Supreme Court,[126] which eventually rejected the petitioner's arguments that the 2010 law's definition of spoofing is 'unconstitutionally vague' and confirmed a 3-year prison sentence. Another pending spoofing case involves Mr. Navinder Sarao, who faced criminal and civil spoofing charges in the US in relation to the aforementioned 2010 'flash crash',[127] to which Mr. Sarao pled guilty. The outcome of this case is highly anticipated in both legal and algorithmic trader circles. However, the verdict in the *US v. Coscia* case was a clear indication that such practices are not being tolerated. Combined with the relevant European legal framework of MiFID II/MiFIR these regulations may be useful instruments in the hands of Fraud Squad officers and prosecutors, in order to investigate further cases of market manipulation. On the other hand, they could form a powerful legal/judicial barrier for traders engaging in algorithmic trading or HFAT, to consider adjusting their practices.

An alternative way to approach algorithmic trading is the concept of providing real rights to conducts that belong in the sphere of a virtual world.[128] A virtual world can be an online interactive one, where players create avatars performing various tasks and activities. The interaction between players, their virtual property, creates a virtual economy, which imitates that of the real world. The relations between providers and players are usually regulated via End-User License Agreement (EULA) and the standard terms and conditions of the game hosting site.[129] But what is valid for a gaming community can be also valid for a community of traders who are engaging in real markets via trading venues. Hence, as long as a subscription fee is required, the status of consumer can be evoked by traders as dictated by EU's Directive 93/13 (Unfair Terms in Consumer Contracts) and Directive 2005/29 (Unfair Commercial Practices Directive), in order for them to benefit, for example, from bringing the case in their area of jurisdiction.[130] In instances of illegal handling of virtual rights, contractual relationship can be examined to establish contractual liability between involved parties. The introduc-tion of digital signatures, security protocols and procedures either as regulations or technical standards have been proposed in order to safeguard those rights and the benefits and properties of their owners.[131]

---

[124] *US v. Coscia*, 14-cr-00551.

[125] See 7 USC§ 6c (a) (5) (C) for a definition of spoofing: 'bidding or offering with the intent to cancel the bid or offer before execution'.

[126] *Coscia v. US*, 17-1099.

[127] *US v. Sarao*, 1:15-cr-00075.

[128] See, e.g., Yannopoulos (2012b).

[129] See Yannopoulos (2012b), p. 2.

[130] See Yannopoulos (2012b), p. 3.

## 3.7 General Data Protection Regulation

Rapid technological developments in telecommunications, globalization of services, the rise of internet and the widespread use of mobile electronic devices have triggered, among others, the need to modify Directive 2002/58/EC concerning the processing of personal data and the protection of privacy in the electronic communications sector (Directive on privacy and electronic communications). After years of considerations and consultation with relevant stakeholders, the EU adopted on 27 April 2016 its General Data Protection Regulation (GDPR),[132] which became directly applicable on 25 May 2018 in all EU MS, without the necessity for additional national legislation. Moreover, the EC has established an expert group on Regulation (EU) 2016/679 to 'clarify how Member States' existing and future legislation will ensure effective and uniform application of the Regulation'.[133] Among other changes, GDPR alters the rules for 'profiling', i.e. the use of computerized data analysis to generate user profiles. As will be discussed below, profiling is only allowed on an opt-in basis and special consent is required to automatically generate the profile of a data subject.

The regulation contains well-known principles of pre-existing data protection law, such as 'consent', the 'accountability principle' and 'privacy by design'.[134] Under the GDPR, both business customers and data controllers have certain legal obligations[135]. Despite the fact that the basic data protection principles remain untouched, the structures and procedures for the protection of the data subjects change. For instance, the consent of the data subject in Art. 4(11) is far more detailed:

> ''consent' of the data subject means any freely given, specific, informed and unambiguous indication of the data subject's wishes by which he or she, by a statement or by a clear affirmative action, signifies agreement to the processing of personal data relating to him or her;'

---

[131] See Yannopoulos (2012b), p. 7.

[132] Regulation (EU) 2016/679.

[133] The European Commission has established a relevant register that includes information on expert groups; for more details on the work of the expert group for GDPR see EC register (2016).

[134] See also Cavoukian (2010).

[135] It is not the purpose of this book to provide a full analysis of the GDPR. Nevertheless, some of its key concepts are to be highlighted.

Compared to previous legislation new key elements insist that the indication of consent must be 'unambiguous and involve clear affirmative action' (opt-in).[136] Art. 7 provides several additional conditions for consent, specifically on keeping records of consent, clarity of consent and the right to withdraw consent. In addition, Art. 8 of the GDPR contains new provisions regarding children's consent for online services. As noted by UK's Information Commissioner's Office (ICO), the opt-in approach pursued through the GDPR is going to positively affect individuals' rights in a sense that existing rights are going to be further strengthened when data processing happens on a consent-basis.[137] The ICO also mentions two related examples of such rights, the 'right to be forgotten'[138] and the 'right to data portability'.[139] At this point, it is interesting to investigate what happens to existing data that have been collected on an opt-out basis. Marketing[140] under the GDPR is regulated like any other data processing activity and Art. 6 implies that the controller or a third party needs to demonstrate that it has a lawful basis to conduct such activities. Moreover, the data subject needs to be aware of who is collecting the data, for which purpose and by whom they are to be further processed. Otherwise, data collection and any processing actions are not lawful. In relation to direct electronic marketing,[141] a widely used business practice, it is noted that:

> '[i]f marketing is already lawfully conducted on an opt-out basis, the GDPR is unlikely to change this (…). If conducted on an opt-in basis, then further review and risk assessment may be needed.'[142]

This assessment comes to the result that businesses which previously went beyond legal necessity in order to protect their client's rights using an opt-in approach, will find themselves in the situation to refresh those consents for GDPR compliance, hence risking to compromise valuable marketing data in case several data subjects will not re-consent.[143] Art. 25 of the GDPR mentions two powerful dimensions in data protection using the terms 'by design' and 'by default'.[144] A close look at Art. 25 is necessary to further study the effects and implications of the provisions therein. [145]

---

[136] See UK ICO (2017), p. 5.

[137] See UK ICO (2017). p. 7.

[138] On the 'right to be forgotten' see also *supra* Section 3.4.

[139] The right to data portability is mentioned in Art. 20 GDPR.

[140] Any form of marketing is concerned, e.g. postal, phone, e-mail, SMS etc.

[141] In the EU electronic marketing activities are also regulated by Directive 2002/58/EC on the processing of personal data and the protection of privacy in the electronic communications sector.

[142] See Lee (2017).

[143] See Lee (2017).

[144] Art. 25(1) GDPR mentions data protection 'by design'. Data protection 'by default' is mentioned in Article 25(2) GDPR. Article 25(3) GDPR enters on a voluntary basis a certification mechanism in place.

Data protection 'by design' in not an original term coined by the GDPR drafters. It means that '(…) privacy is embedded into the design and architecture of IT systems and business practices'.[146] Moreover, already in 2014, the European Union Agency for Network and Information Security (ENISA) produced a relevant report to bridge the gap between the pre-GDPR legal framework and available technologies.[147] Its goal is to motivate all stakeholders towards more privacy-friendly systems and services design and development. When it comes to implementation, several potentially limiting conditions such as 'state of the art' and 'cost of implementation' need to be taken into consideration. These exact two conditions are also to be found in Art. 17(1) of the previous Data Protection Directive 95/46/EC.

Data protection 'by default' means that privacy settings are originally implemented and built into the system. Hence, there is no necessity for a data subject to proceed to any actions to protect its privacy. The existence of predefined parameters, usually presented as initial pre-settings in the Graphical User Interface (GUI), does not mean that changes are not possible. However, most users do not change pre-settings, acting in good faith or these are skipped due to time considerations. Data protection 'by Default' implies the use of the 'Principle of Least Surprise' (or Astonishment) in software development, which states that the result of an operation should be 'obvious, consistent, and predictable'.[148] Data protection 'by Default' contains less potentially limiting conditions compared to data protection 'by design' making it a potentially more powerful principle. Instead, it is related to the 'purpose limitation' principle from Art. 5(1) GDPR.[149] According to the GDPR, companies or organizations that deal with EU personal data need to decide on the appointment of a Data Protection Officer (DPO).[150] This is rather linked to the volume of sensitive personal information such an entity may process. For instance, a company that routinely handles large volumes of personal data should be more inclined towards appointing a DPO. In general, a core requirement that arises from Art. 25 is that companies or organizations must implement mechanisms for data protection, such as security or privacy checks, in all layers of operation, such as in their management, workings procedures and the provision of products or services.

---

[145] Art. 25 GDPR Data protection by design and by default:

1. Taking into account the state of the art, the cost of implementation and the nature, scope, context and purposes of processing as well as the risks of varying likelihood and severity for rights and freedoms of natural persons posed by the processing, the controller shall, both at the time of the determination of the means for processing and at the time of the processing itself, implement appropriate technical and organisational measures, such as pseudonymisation, which are designed to implement data-protection principles, such as data minimization, in an effective manner and to integrate the necessary safeguards into the processing in order to meet the requirements of this Regulation and protect the rights of data subjects.

2. The controller shall implement appropriate technical and organizational measures for ensuring that, by default, only personal data which are necessary for each specific purpose of the processing are processed. That obligation applies to the amount of personal data collected, the extent of their processing, the period of their storage and their accessibility. In particular, such measures shall ensure that by default personal data are not made accessible without the individual's intervention to an indefinite number of natural persons.

3. An approved certification mechanism pursuant to Art. 42 may be used as an element to demonstrate compliance with the requirements set out in paragraphs 1 and 2 of this article.

[146] See Cavoukian (2010), p. 3.
[147] See Danezis et al. (2015).

However, Art. 25 provides DPOs with relatively little technical guidance on how to accomplish such a widespread and complex task. There are two important terms that are mentioned here, 'pseudonymisation' and 'data minimisation'. In addition, the GDPR mentions the term 'encryption' elsewhere.[151] The technical term 'pseudonymisation' has here its first appearance in EU law. This new term refers to the processing of personal data in a way that the result cannot be attributed to a specific data subject without the use of further related information.[152] Those are to be stored separately, using appropriate technical and security measures. Another significant requirement of Art. 25 is the so-called 'data minimisation', which results in the processing of only a necessary amount of data for a specific purpose, although, in practical terms, this could be a difficult requirement to achieve.[153] En-cryption is to be considered as a way to achieve data security, and hence privacy, by complicating and, ultimately, blocking data access to unauthorized third parties and data processors. Also, data subjects are to be allowed to access and alter their privacy settings and information with the same ease as giving their consent to use and process their data, in analogy to Art. 7(3).[154] A further critical examination of Art. 25 GDPR reveals several shortcomings that include:[155]

---

[148] See Gunderloy (2005), p. 128.

[149] See relevant presentation by Hansen (2017), data protection commissioner in Schleswig-Holstein, Germany.

[150] See Art. 39 GDPR for the responsibilities and tasks of a DPO. A DPO should not be a data controller as well and vice versa.

[151] See, for instance, Art. 6, 32 and 34 GDPR.

[152] The anonymization of data prior to the release of legal documents has been used previously as a balanced solution between the rights of access to information and protection of personal data. Approaches greatly differ from country to country; a relevant analysis for the domain of legal databases is offered by Yannopoulos (2012a); a recent example of pseudonymization is offered by Decisions 1/2017 and 4/2017 of the Hellenic Data Authority in the case of introduction of the new electronic ticket for public transportation in Athens. 'Hashing' has been applied for the issuance of each new personalized card; hence, no documented journey can be traced back to the cardholder but to a hash value.

[153] From a technical point of view, overcollection or over processing of data cannot be easily proven as these are relative terms (e.g. overcollection in relation to what?). There is no such thing as an optimal algorithm since algorithms are constantly developed and optimized.

- 'fuzzy legalese',
- insufficient clarity of parameters and methodologies to achieve its goals,
- lack of salient and strong incentives and
- missing communication link to the designers and developers of information systems.

These shortcomings demonstrate how difficult it is to impose regulation on a highly dynamic technology sector.

Summarizing, the provisions of Art. 25 have serious implications to a series of parameters of data processing, while access to the original and processed data by third parties is generally not permitted:[156]

- amount of data collected,
- extent to which the data is processed,
- storage duration,
- data accessibility.

DPOs need to bear in mind the above, in order to comply with the special GDPR requirements from Art. 25. Other key changes that the GDPR introduces include stricter rules to report security breaches (Art. 33(1))[157], as well as the accountability principle, according to which there is a controller responsibility to 'demonstrate compliance' with personal data processing rules (Art. 5(2)). From the above, it becomes obvious that the GDPR constitutes a decisive evolution to the pre-existing EU legal framework that provides public authorities with regulatory guidelines, as well as a certain level of guidance to private sector and individual stakeholders. Nevertheless, the implementation of such a comprehensive privacy framework needs to be monitored closely, both at the national, as well as the EU level, as it is likely to cause negative implications in indirectly related fields, such as in advertising.[158] Certainly, the first judicial judgements on implementation of the GDPR are awaited with great interest, as they are expected to provide clarity to some of the ambiguous concepts represented therein. Moreover, and inline with other scholars, we expect a gradual establishment of relevant national and international regulatory bodies or agencies.[159]

---

[154] See Art. 7(3): 'The data subject shall have the right to withdraw his or her consent at any time. (…) It shall be as easy to withdraw as to give consent'.

[155] See Bygrave (2017), p. 119.

[156] As previously commented, access to the data may be granted with the consent of the data subject(s).

[157] A legal obligation is introduced to notify the supervisory authority of a personal data breach within 72 hours of knowing about it.

# References

Airbnb. 2019a. Fast facts. https://press.atairbnb.com/fast-facts/. Accessed 4 April 2019.

Airbnb. 2019b. Next steps in Amsterdam. https://www.airbnbcitizen.com/next-steps-in-amsterdam/. Accessed 22 April 2019.

BGH. 2018. Bundesgerichtshof: Angebot des Werbeblockers AdBlock Plus nicht unlauter. Press announcement (19 April 2018). http://juris.bundesgerichtshof.de/cgi-bin/rechtsprechung/document.py?Gericht=bgh&Art=pm&Datum=2018&Sort=3&nr=82856&pos=0&anz=78. Accessed 30 May 2018.

Bygrave, Lee A. 2017. Data Protection by Design and by Default: Deciphering the EU's Legislative Requirements. *Oslo Law Review* 4(2): 105-120. https://doi.org/10.18261/issn.2387-3299-2017-02-03.

Cansoy, Mehmet, and Juliet Schor. 2017. Who Gets to Share in the "Sharing Economy"? Racial Discrimination in Participation, Pricing and Ratings on Airbnb. Working paper (unpublished). Boston University.

Cavoukian, Ann. 2010. Privacy by Design - The 7 Foundational Principles - Implementation and Mapping of Fair Information Practices, Information and Privacy Commissioner of Ontario, 1-10. Canada. http://www.ontla.on.ca/library/repository/mon/24005/301946.pdf. Accessed 29 May 2018.

Cea-Esteruelas, Nereida. 2013. Cybermedia economics: revenue model and sources of financing. *El profesional de la información* 22(4): 353-361. https://doi.org/10.3145/epi.2013.jul.12

Cohen, Amanda. 2004. Surveying the Microsoft Antitrust Universe. *Berkeley Technology Law Journal 19(1)*:333-364. https://doi.org/10.15779/Z38MH4K.

Cunningham-Parmeter, Keith. 2016. From Amazon to Uber: Defining employment in the modern economy. *Boston University Law Review* 96: 1673-1728.

Danezis, George, Stefan Schiffner, Marit Hansen, Rodica Tirtea, Josep Domingo-Ferrer, Daniel Le Métayer, and Jaap-Henk Hoepman. 2015. *Privacy and Data Protection by Design – from policy to engineering.* Heraklion: ENISA. https://doi.org/10.2824/38623.

De Carbonnel, Alissa. 2017. Volkswagen to offer EU diesel car owners extended warranty but no money back: EC. *Reuters*, June 14. https://www.reuters.com/article/us-volkswagen-emissions-idUSKBN19527H. Accessed 19 June 2018.

Dörting, Thorsten, Matthias Streitz, and Jörn Sucher. 2017. So finanziert sich SPIEGEL ONLINE. *Spiegel Online*, August 16. http://www.spiegel.de/extra/werbung-plus-daily-so-finanziert-sich-spiegel-online-a-1162309.html. Accessed 30 May 2018.

Dredge, Diane, Szilvia Gyimóthy, Andreas Birkbak, Torben Elgaard Jensen, and Anders Koed Madsen. 2016. The impact of regulatory approaches targeting collaborative economy in the tourism accommodation sector: Barcelona, Berlin, Amsterdam and Paris. Impulse Paper No 9 prepared for the European Commission DG GROWTH. Copenhagen: Aalborg University. https://ec.europa.eu/docsroom/documents/19121/attachments/1/translations/en/renditions/native. Accessed 22 April 2019.

---

[158] See, e.g., Goldfarb and Tucker (2011), who investigated the impact of privacy regulation on online advertising in the EU; empirical evidence is provided that privacy regulation can reduce the effectiveness of advertising.

[159] The development of regulatory bodies is presented in Chapter 4.

EC register. 2016. Commission expert group on the Regulation (EU) 2016/679 and Directive (EU) 2016/680(E03461).http://ec.europa.eu/transparency/regexpert/index.cfm?do=groupDetail.groupDetail&groupID=3461. Accessed 29 May 2019.

EC. 2018. Press release (20 September 2018). EU consumer rules: Airbnb commits to complying with European Commission and EU consumer authorities' demands. http://europa.eu/rapid/press-release_IP-18-5809_en.htm. Accessed 22 April 2019.

Economides, Nicholas. 2001. The Microsoft antitrust case. *Journal of Industry, Competition and Trade* 1 (1): 71-79. https://doi.org/10.1023/A:1011576827599.

Economides, Nicholas. The Microsoft Antitrust Case. *NYU Law and Economics Working Paper Series* 01-003. http://dx.doi.org/10.2139/ssrn.253083.

Edelman, Benjamin G and Damien Geradin. 2016. Efficiencies and regulatory shortcuts: How should we regulate companies like Airbnb and Uber. *Stanford Technology Law Review* 19 (2): 293-328.

Edelman, Benjamin, and Wesley Brandi. 2015. Risk, Information, and Incentives in Online Affiliate Marketing. *Journal of Marketing Research* 52(1): 1–12. https://doi.org/10.1509/jmr.13.0472.

ESMA. 2014a. Discussion paper MiFID II/MiFIR. 2014/548. https://www.esma.europa.eu/sites/default/files/library/2015/11/2014-548_discussion_paper_mifid-mifir.pdf. Accessed 28 May 2019.

ESMA. 2014b. Consultation paper MiFID II/MiFIR. 2014/549. https://www.esma.europa.eu/sites/default/files/library/2015/11/2014-549_-_consultation_paper_mifid_ii_-_mifir.pdf. Accessed 28 May 2019.

European Commission (2017). Consumer Authorities and the European Commission urge Volkswagen to finalise repairs of all cars affected by emissions scandal. Press release (7 September 2017). http://europa.eu/rapid/press-release_IP-17-3102_en.htm. Accessed 19 June 2018.

European Parliament. 2015. Setting up a Committee of Inquiry into emission measurements in the automotive sector, its powers, numerical strength and term of office (2015/3037(RSO)) (17 December 2015). http://www.europarl.europa.eu/sides/getDoc.do?type=TA&reference=P8-TA-2015-0462&language=en. Accessed 18 June 2018.

European Parliament. 2017. Report on the inquiry into emission measurements in the automotive sector (2016/2215(INI)). http://www.europarl.europa.eu/doceo/document/A-8-2017-0049_EN.html. Accessed 18 June 2018.

Frantziou, Eleni. 2014. Further Developments in the Right to be Forgotten: The European Court of Justice's Judgment in Case C-131/12, Google Spain, SL, Google Inc v Agencia Espanola de Proteccion de Datos. *Human Rights Law Review* 14 (4): 761-777. https://doi.org/10.1093/hrlr/ngu033.

Gitter, Donna M. 2004. Strong Medicine for Competition Ills: the Judgement of the European Court of Justice in the IMS Health action and its implications for Microsoft Corporation. *Duke Journal of Comparative & International Law* 15(1): 153-192.

Goldfarb, Avi. and Catherine E. Tucker. 2011. Privacy regulation and online advertising, *Management science* 57.1. https://doi.org/10.1287/mnsc.1100.1246. Accessed 29 May 2019.

Goldsmith, Jack L., and Alan O. Sykes. 2001. The Internet and the dormant commerce clause. *The Yale Law Journa*l 110(5): 785-828.

Google. 2019. Privacy & Terms. https://policies.google.com/faq?hl=en. Accessed 4 April 2019.

Gunderloy, Mike. 2005. *Developer to designer: GUI design for the busy developer.* Alameda: SYBEX.

Hansen, Marit. 2017. Data Protection by Default – Requirements, Solutions, Questions. Presentation at the IPEN Workshop. June 9. Vienna. https://edps.europa.eu/sites/edp/files/publication/17-06-09_marit_hansen-dataprotectionbydefault_ipen-workshop_vienna_hansen_en.pdf. Accessed 29 May 2019.

Harty, Ronan P., and Gregory G. Wrobel, eds. 2003. *2002 Annual Review of Antitrust Law Developments*. Chicago: American Bar Association.

Interian, Johanna. 2016. Up in the air: Harmonizing the sharing economy through Airbnb regulations, *Boston College International and Comparative Law Review* 39 (1): 129-161.

Juul, Maria. 2016. Lawsuits triggered by the Volkswagen emissions case. European Parliamentary Research Service. PE 583.793. Brussels: European Parliament. http://www.europarl.europa.eu/RegData/etudes/BRIE/2016/583793/EPRS_BRI(2016)583793_EN.pdf. Accessed 16 June 2018.

Klein, Joel, and Preeta Bansal. 1996. International Antitrust Enforcement in the Computer Industry. *Villanova Law Review* 41(1): 173-192.

Lee, Phil. 2017. Re-consenting to marketing under GDPR?. https://privacylawblog.fieldfisher.com/2017/re-consenting-to-marketing-under-gdpr. Accessed 29 May 2019.

Leon, Pedro, Blase Ur, Richard Shay, Yang Wang, Rebecca Balebako, and Lorrie Cranor. 2012. Why Johnny can't opt out: a usability evaluation of tools to limit online behavioral advertising. In *Proceedings of the SIGCHI Conference on Human Factors in Computing Systems* (CHI '12), 589-598. New York: ACM. https://doi.org/10.1145/2207676.2207759

Liberman, Michael. 1995. Overreaching Provisions in Software License Agreements. *Richmond Journal of Law & Technology* 1(1): 4.

Margaritis, Konstantinos. 2018. The Role of Judicial Dialogue between the CJEU and the ECtHR in the Formulation of the Right of Privacy. In *The Right to be Forgotten in Europe and Beyond/Le droit à l'oubli en Europe et au-delà*, eds. Olivia Tambou, and Sam Bourton, 98-99. Luxembourg: Blogdroiteuropéen.

Marsden, Chris, and Trisha Meyer. 2019. Regulating disinformation with artificial Intelligence. Brussels: European Parliament. https://doi.org/10.2861/003689.

Matsui, Shigenori. 2019. Is Law Killing the Development of New Technologies: Uber and Airbnb in Japan. *Boston University Journal of Science and Technology Law* 25: 100-144.

Nordemann, Jan Bernd. 2018. Liability of Online Service Providers for Copyrighted Content – Regulatory Action Needed? European Parliament, IP/A/IMCO/2017-08. http://www.europarl.europa.eu/RegData/etudes/IDAN/2017/614207/IPOL_IDA(2017)614207_EN.pdf. Accessed 28 May 2019.

Norton Rose Fulbright. 2014. High frequency and algorithmic trading obligations. MiFID II / MiFIR series. http://www.nortonrosefulbright.com/knowledge/publications/115236/mifid-ii-mifir-series. Accessed 28 May 2019

Rivet, Myriam. 2018. Paris asks court to fine Airbnb over unregistered listings. *Reuters*, April 12. https://www.reuters.com/article/france-airbnb/paris-asks-court-to-fine-airbnb-over-unregistered-listings-idUSL8N1RP1LQ. Accessed 19 June 2018.

Rosen, Jeffrey. 2011. The right to be forgotten, *Stanford Law Review Online* 64: 88-92.

UK ICO. 2017. Consultation: GDPR consent guidance. https://ico.org.uk/media/about-the-ico/consultations/2013551/draft-gdpr-consent-guidance-for-consultation-201703.pdf. Accessed 29 May 2019.

Ulin, Jeff. 2013. *The Business of Media Distribution: Monetizing Film, TV and Video Content in an Online World*. New York: Routledge.

US District Court (2018). Volkswagen "Clean Diesel" MDL. https://www.cand.uscourts.gov/crb/vwmdl. Accessed 18 June 2018.

US DoJ docket for commodities fraud. https://www.justice.gov/criminal-fraud/commodities-fraud. Accessed 29 May 2019.

US DoJ. 2016. Press release, January 4. https://www.justice.gov/opa/pr/united-states-files-complaint-against-volkswagen-audi-and-porsche-alleged-clean-air-act. Accessed 18 June 2018.

US DoJ. 2017. Press release, January 11. https://www.justice.gov/opa/pr/volkswagen-ag-agrees-plead-guilty-and-pay-43-billion-criminal-and-civil-penalties-six. Accessed 18 June 2018.

US EPA. 2015a. Notice of Violation (18 September 2015). https://www.epa.gov/sites/production/files/2015-10/documents/vw-nov-caa-09-18-15.pdf. Accessed 16 June 2018.

US EPA 2015b. Notice of Violation (2 November 2015). https://www.epa.gov/sites/production/files/2015-11/documents/vw-nov-2015-11-02.pdf. Accessed 18 June 2018.

Wachter, Sandra. 2019. Data protection in the age of big data. *Nature Electronics* 2:6-7. https://doi.org/10.1038/s41928-018-0193-y. Accessed 29 May 2019.

Wojdynski, Bartosz, and Nathaniel J. Evans. 2015. Going Native: Effects of Disclosure Position and Language on the Recognition and Evaluation of Online Native Advertising. *Journal of Advertising* 45(2): 157-168. http://dx.doi.org/10.1080/00913367.2015.1115380

WP29. 2018. Guidelines on Automated individual decision-making and Profiling for the purposes of Regulation 2016/679 (wp251rev.01). http://ec.europa.eu/newsroom/article29/item-detail.cfm?item_id=612053. Accessed 7 June 2018.

Yannopoulos, Giorgos. 2012a. M., A. kai L. katá S. I anonymopoíisi ton váseon nomikón dedoménon. *Díkaio Méson enimérosis kai Epikoinonías* 1: 21-27.

Yannopoulos. Georgios. 2012b. Real Rights in Virtual Worlds and Virtual Rights in a Real World. 5th International Conference on Information Law and Ethics. June 29-30. Corfu. https://doi.org/10.6084/m9.figshare.8262299.v1.

Yannopoulos, Giorgos. 2013. I efthýni ton michanismón anazítisis gia tis ypiresíes ypódeixis (suggest) kai aftómatis symplírosis (autocomplete). Scholiasmós tis apófasis Monomeloús Protodikeíou Athinón 11339/2012. *Díkaio Méson enimérosis kai Epikoinonías* 2: 168-171.

# 4 Development of Regulatory Bodies

**Fotios Fitsilis**

**Abstract**

The development of national, supranational or global regulatory bodies for advanced algorithms is essential. In this chapter, a comparison of several approaches is attempted. Among others, the first steps towards the formation of regulatory bodies is explained. Moreover, we present a set of modes for classification or regulatory activities. Finally, the role of parliamentary institutions is highlighted and the idea of an 'algorithmic monitor' based on crowdsourcing is proposed.

Keywords: regulatory body, parliamentary research service, post legislative scrutiny, ENISA, algorithmic monitor.

Should advanced algorithms continue to advance, a structured approach to regulate them should be adopted. In this chapter, the development of regulatory bodies shall be discussed and the pros and cons of different regulatory concepts will be presented, including issues of and opportunities for self-regulation, as well as technical considerations.

Several forms of oversight institutions with regards to algorithms have already been established.[1] In a resolution on civil law rules on robotics the EP expressed its position for the establishment of a '(…) European Agency for robotics and artificial intelligence in order to provide the technical, ethical and regulatory expertise needed to support the relevant public actors (…)'.[2] At the same time, the Committee tried to approximate the potential operational framework for this prospective regulatory agency by mentioning a list of possible duties:[3]

- Cross-sectorial and multidisciplinary monitoring of robotics-based applications,
- Identification of best practice,
- Recommendation of regulatory measures,

---

[1] A number of references to different proposals for algorithmic regulations, such as an AI watchdog, a Machine Learning Commission, a US FDA (Food and Drug Administration) for Algorithms etc. is included in Andrews (2017), pp. 10-11.

[2] See European Parliament (2017), p. 10, §16.

[3] See European Parliament (2017), p. 10, §17.

- Definition of new principles and
- Addressing potential consumer protection issues and systematic challenges.

The interesting point here is that the proposed agency is not a regulatory body in itself. Instead, it would provide public authorities ('public actors') at all levels, i.e. the European Union and EU member states, with an ethical code of conduct and the necessary expertise in order for the latter to proceed to regulatory measures, such as recommendations for the implementation of 'kill switches' in software design.[4] The resolution and the related report have stirred up considerable interest triggering a public consultation with the aim to encourage citizens and stakeholders to share their thoughts and considerations in the fields of robotics and AI.[5] However, DG Connect of the European Commission could be mandated to fulfil this task, instead of EU establishing another costly and quite probably, inefficient agency. Alternatively, an existing agency such as the European Union Agency for Network and Information Security (ENISA) could be tasked with the same mandate, as shall be analyzed below. If not, then there would surely be overlaps of competencies between all three.

A preliminary evaluation of the results of the consultation shows that most respondents have positive attitudes towards robotics and AI, while a large majority expresses the need for public regulation in the area. Interestingly enough, the respondents also specify that regulation should be conducted at EU and/or international level. Some responders go a step further indicating the nature of this body, i.e. the Center for Data Innovation suggested the establishment of a new directorate in the EC for the support of technological advancements in the fields of robotics and AI without focusing on ethical or regulatory issues.[6] In its response to EP's resolution, the EC admitted its intention to investigate over time several aspects and the regulatory dimension of the issue.[7]

In the US, a broad discussion over regulatory matters is also taking place. The regulatory approach is described as '(…) sector-specific, with oversight by a variety of agencies'.[8] Scherer proposed the creation of a federal regulatory regime for AI under a new legislation he calls Artificial Intelligence Development Act (AIDA) by establishing a tort system approach.[9] The proposed federal agency would consist of two components for policymaking and certification, respectively. The establishment of an AI certification process would require companies and de-velopers to seek certification before bringing an AI product or service onto the market.[10] The establishment of a 'Federal Robotics Commission' –as a part of an Einsteinian thought experiment– has also been discussed elsewhere.[11] However, not everybody is in favor of regulatory bodies and there are also voices worrying they could result in hamstringing innovation[12], i.e. with too much red tape. The basic logic behind this position is that the underlying risk analysis is often inadequate, exaggerating risks and downgrading real benefits.

---

[4] See Hasselbalch (2017).

[5] The public consultation has been launched by the European Parliament's Committee on Legal Affairs (JURI) in cooperation with the EPRS via the EP Committee web space; The relevant document stack and first results are presented in JURI (2017), while the summary report is highly anticipated.

[6] See Castro et al. (2017), p. 18.

[7] See Ponce Del Castillo (2017), p. 3.

[8] See Stone et al. (2016), p. 44.

[9] See Scherer (2015), p. 394.

There have been suggestions to create a 'trusted third party' to scrutinize decisions of automated decision-making systems, thus providing an oversight mechanism for the application of advanced algorithms.[13] Such a suggestion would make sense if implemented on the national level; the sheer amount of complaints and the range of applications would constitute a similar super-national authority dysfunctional. Alternatively, a European regulator to audit algorithms - prior to their deployment - is proposed therein.[14]

China is actively challenging US leadership in AI through an aggressive five-year plan (2016-2020). At the heart of China's science and technology policy there is the National Science, Technology and Education Leading Small Group, which is headed by The Premier of the State Council.[15] He has identified five distinct agencies responsible for the development and implementation of central government policies in AI, while there are also other centralized agencies responsible for sectoral and industry-specific regulation.[16]

In order to structure and discuss issues of regulation, a classification in different modes may be attempted.[17] For instance, depending on the timing of regulation (timing mode), one may distinguish between ex-ante, e.g. via administrative decisions or legislation, and ex-post regulation. A classic ex-post regulatory approach would be regulation by judicial decisions. Judicial rulings regulate by definition after a case has developed. In certain disputes, such as in the Microsoft cases, several years of investigations by authorities may precede. Nevertheless, in this early stage of the development of advanced algorithms, we consider ex-ante regulation far more difficult, since regulatory capacity needs first to be built-up in order to

---

[10] See Scherer (2015), p. 395.

[11] See Calo (2014).

[12] See relevant blog post by O'Sullivan (2017).

[13] See Wachter et al. (2017), p. 98.

[14] See Wachter et al. (2017).

[15] See He (2017), p. 4.

[16] See He (2017).

[17] Five regulatory classes, i.e. 'modes' have been proposed: intervention mode, hierarchical mode, natural mode, type mode and timing mode; see also Chapter 1 for definitions and more details; the type and nature modes will be highlighted in Chapter 5, during analysis of the case studies.

keep up with growing innovation in an increasingly complex interdisciplinary field.[18]

A further distinction would concern the hierarchy of regulation. Fig. 4.1 presents a proposal for the hierarchical classification of regulation for advanced algorithms. This may be also linked to the positioning of the regulatory body in the chain of multi-level governance. In this regard, a national body, such as an Independent State Body (ISB), governmental agency or unit, a supranational body, such as an EU agency, or a global intergovernmental body, could be considered as appropriate candidates. Depending on its positioning, a regulatory body would have a different set of legal instruments at its disposal. These may include technology standards, pertinent legal frameworks and court rulings of all relevant instances. Standards, such as the Akoma Ntoso V1.0 OASIS standard for legal documents or the European Case Law Identifier (ECLI) for citing judgments from European and national courts,[19] may not be used *per se* in the regulatory process. Usually, they first need to be adopted or endorsed by administrative decisions, or transposed by regulatory acts in the national or supranational legal orders. We then speak of *de jure* standards. This can be attributed to the fact that several standards are independently developed by standardization bodies, such as the nongovernmental International Organization for Standardization (ISO), or private companies.[20] In a profound analogy to the sources of international law, additional grounds for reaching a regulatory decision may include scholarly opinions, customary practices and general principles of law.[21]

Another, more liberal, possibility is for a complimentary system of self-regulation amongst private sector actors, reporting and answerable to the established or mandated government body.[22] In the social media regime, where Facebook indisputably enjoys a position of power, an interesting new approach arises. According to the company, there are plans to create an 'independent body' of experts, who will take responsibility for content-related decisions.[23] Facebook names this body 'Oversight Board' and underlines its independence, a claim that will surely be challenged given that its members are both going to be remunerated and supported, in terms of full-time staff, by Facebook itself. Therefore, the idea to privatize regulatory action in critical domains, such as private communications and public policy discourse, is not likely to find many supporters among regulators, the European ones in particular.

---

[18] See also the relevant Quora contribution by Zhao (2017).

[19] See Akoma Ntoso (2018) and European Council (2011), respectively.

[20] When it comes to standardization, Marx (2017, p. 26) argues that the distinction between public and private is blurring, as both sectors increasingly co-regulate.

[21] According to the Statute of the International Court of Justice (ICJ), Art. 38 (1); on the other hand, it is highly questionable whether a regulatory body will ever be able to decide a case *ex aequo et bono*.

[22] This would belong to the 'intervention mode'; more on the self-regulatory regime in Finck (2017), as well as in Trubek and Trubek (2007).

[23] See Clegg (2019) and the draft charter therein.

From the relevant discussion it becomes clear that there is no consensus of how regulation should look like or, at least in some cases, what is to be regulated. However, should societies come to an agreement on the necessity to regulate advanced algorithms, and despite the level and the extent of regulation, one needs to think about the timing of regulation. Our understanding is that regulation is inevitable and needs to be applied, sooner rather than later, in the development cycle of technologies related to advanced algorithms, based on the previously mentioned consideration that regulators need to be both educated and trained in the early stages of this rapidly developing field.

Currently, possibly with the exception of China, most states and supranational state unions, such as the European Union, mainly apply ex-post regulation of advanced algorithms via judicial decisions of all instances. Our analysis finds that a centralized regulatory model is favored among most scholars and stakeholders. In the EU, this could take the form of an EU agency, as mentioned above, but a hybrid approach of a central agency with national offices could also be considered given the scattered nature of legal order in each EU member state.[24] ENISA, given its existing status as an EU agency and its mandate could be an interesting incubator for developing regulatory expertise on advanced algorithms, particularly when considering that it has conducted related work.[25] Thus, a future regulatory body could be an ENISA spin-off that incorporates both existing technical know-how and advisory expertise.

On the national level, this body would be complemented by institutions in the form of existing or, where not present, new ISBs. These institutions would be the eyes and ears of the competent EU agency at the member state level. Not only would they be responsible for transposing EU guidelines and regulations into the national legal framework, but they could also be used as a local observatory for issues of algorithmic relevance. This so-called 'algorithmic monitor' could use the capacity of the crowd, i.e. crowdsourcing,[26] in order to spot early and analyze cases where algorithmic regulation may apply. Fig. 4.2 depicts a principal design of an algorithmic monitor. It relies on the 'wisdom of the crowd' to collect information on specific algorithms or areas that may require regulation. These raw pro-posals need to undergo a processing step, which may include filtering, prioritization and deeper analysis. The output of the aforementioned process would then be made available for study by a supervisory body or regulatory agency, which by definition have limited capacity to equally handle any incoming proposal or complaint.[27]

---

[24] See, for instance, the organization and operation of both the European Patent Office (EPO) and the national patent offices. The central instance, the EPO, has taken the form of an intergovernmental organization. Another option could be the an analogon to ENISA, an agency for network and information security, that works together with EU member states and the private sector to deliver strategic advice and solutions.

[25] See relevant ENISA report on privacy and data protection by Danezis et al. (2015).

[26] Orozco (2016) analyzes, among others, the effects of crowdsourcing on legal, regulatory and policy issues.

Parliaments are democracy's supreme institutions. As algorithms play an increasingly important role in people's lives, their role in algorithmic regulation needs to be discussed. As a matter of fact, we consider the parliament's involvement so significant that it deserves a dedicated study. One may define at least two major situations where a national parliament[28] may express opinion, guide or directly regulate in cases that involve advanced algorithms. Again, one must differentiate between the ex-ante and the ex-post approach. Here, the point of reference would be formed by the time a specific algorithm or a special category of algorithms[29] hits the market or becomes widely operational outside the developer's confined testbed.

Most times, the ex-ante regulation would equal the standard legislative approach. This is considered as the most demanding stage, as legal drafts are usually prepared by the government and then submitted to the parliament for subsequent amendments and discussion. National parliaments retain the right to submit proposals of law. However, in the case of complex topics such as advanced algorithms, it may be doubted whether or not parliaments own the internal capacity to study in-depth and subsequently formulate adequate legal provisions. Beyond legal elaboration, after a law has passed, one is entering the regime of parliamentary control. Modern parliaments apply new methods for screening implementation of a law known as Post-Legislative Scrutiny (PLS).[30] Nevertheless, to date, most parliaments lack sufficient capacity to systematically follow up on the implementation of passed legislation. Instead, the traditional system of –written or oral– questions is used by parliamentary groups and MPs to exercise parliamentary control. PLS can be a domain where parliaments may increase their leverage against the executive, particularly when related to the evaluation of regulations related to advanced algorithms.

In the parliamentary context, the ex-post regulatory approach to advanced algorithms would be similar to the one that a parliament uses on several occasions. This would involve the forming of parliamentary committee(s), hereafter committee(s), in order to discuss a topic on algorithmic regulation. The type of the com-mittee also defines its importance within the parliamentary universe.[31] Committees usually have the right to invite external experts, such as academics or consultants, to present their opinion and throw light on the issues discussed, while parliaments may also form research or advisory committees with non-MPs as members.

---

[27] It may also be noted that the algorithmic monitor may be an algorithm on its own, which automatically screens, prioritizes and forwards the most 'significant' regulatory proposals to its human operators.

[28] That supranational parliaments may play a role has been already demonstrated in the case of the European Parliament (2017).

[29] For instance, algorithms that are utilized in HFAT constitute a special category that is regulated by the EU's MiFID II legislative framework.

[30] See De Vrieze and Hasson (2017).

Furthermore, parliaments have the right to discuss in the competent committees or even in the plenary a report submitted by an agency, just like the proposed EU agency on algorithmic regulation, or a national ISB. Undoubtedly, this would constitute the least invasive option, since no new parliamentary bodies would have to be formed and one could rely on existing procedures without the need to change the standing orders. Typically, such discussions result in a resolution that is addressed to the competent Ministry or the Government as a whole, which is then called to transform it to relevant administrative actions, e.g. draft laws or administrative decrees. However, practice has shown that the result of parliamentary action is usually non-tangible and of limited regulatory impact.

Parliamentary Research Services (PaRS) can have a significant role in strengthening the operations and impact of *representative institutions*. This is why most Parliaments have established PaRS and continue to invest considerable resources in their further development.[32] In order to fulfill their role at the highest possible level, PARS are necessary to employ highly skilled researchers that have advanced expertise in a wide range of fields.[33] The work of researchers can be linked to the application of scientific methodology, the following of a code of conduct and, most importantly, the publication of elaborated material.[34] In recent years, an increasing demand for more complex and synthetic information from PaRS can be attested. Advanced algorithms constitute a wide and active field of study and PaRS have the potential to:

> '(…) provide internal and external clients with independent, well-researched, timely, structured and concentrated knowledge products, thus counterbalancing partisan information flows or even governmental superiority in analysis and dissemination of information'.[35]

In order to be able to do so, PaRS clearly need to significantly advance their relevant capacity, mainly in scientific fields, such as big data, data ethics and legal informatics, for example.

---

[31] Parliamentary committees may be formed on a regular or ad-hoc basis. These are also several levels of committees, such as Standing Committees, Permanent Committees, Special Permanent Committees etc., according to their significance in the parliamentary context.

[32] The Inter-Parliamentary Union (IPU) and the International Federation of Library Associations and Institutions (IFLA) have published in 2015 guidelines for PaRS in order to help developing legislatures to establish research services as well as to strengthen existing ones; see IPU and IFLA (2015).

[33] On the researcher role in parliaments see Fitsilis (2018), pp. E-48–E-50.

[34] See Fitsilis (2018).

[35] See Fitsilis and Koutsogiannis (2017), p. 11.

# References

Akoma Ntoso. 2018. Version 1.0 Part 1: XML Vocabulary. OASIS Standard. Edited by Mon-ica Palmirani, Roger Sperberg, Grant Vergottini, and Fabio Vitali. http://docs.oasis-open.org/legaldocml/akn-core/v1.0/os/part1-vocabulary/akn-core-v1.0-os-part1-vocabulary.html. Accessed 27 June 2019.

Andrews, Leighton. 2017. Algorithms, governance and regulation: beyond 'the necessary hashtags'. In *Algorithmic regulation*, eds. Leighton Andrews et al., 7-12. Lodon: Centre for Analysis of Risk and Regulation. http://www.lse.ac.uk/accounting/assets/CARR/documents/D-P/Disspaper85.pdf

Calo, Ryan. 2014. The Case for a Federal Robotics Commission. September 2014. Brookings Institution Center for Technology Innovation. https://www.brookings.edu/wp-content/uploads/2014/09/RoboticsCommissionR2_Calo.pdf. Accessed 20 March 2019.

Castro, Daniel, Nicholas Wallace, and Joshua New. 2017. Response to the European Parliament's Public Consultation on Civil Law Rules on Robotics. Center for Data Innovation, http://www2.datainnovation.org/2017-eu-ai-public-consultation.pdf. Accessed 20 March 2019.

Clegg, Nick. 2019. Charting a Course for an Oversight Board for Content Decisions. January 28. https://newsroom.fb.com/news/2019/01/oversight-board/. Accessed 28 June 2019.

Danezis, George, Stefan Schiffner, Marit Hansen, Rodica Tirtea, Josep Domingo-Ferrer, Daniel Le Métayer, and Jaap-Henk Hoepman. 2015. *Privacy and Data Protection by Design – from policy to engineering*. Heraklion: ENISA. https://doi.org/10.2824/38623.

De Vrieze, Franklin and Victoria Hasson. 2017. Comparative study of practices of Post-Legislative Scrutiny in selected parliaments and the rationale for its place in democracy assistance. Westminster Foundation for Democracy. https://www.wfd.org/wp-content/uploads/2018/07/Comparative-Study-PLS-WEB.pdf. Accessed 21 March 2019.

European Council. 2011. Council conclusions inviting the introduction of the European Case Law Identifier (ECLI) and a minimum set of uniform metadata for case law (2011/C 127/01). https://eur-lex.europa.eu/legal-content/EN/ALL/?uri=CELEX:52011XG0429(01). Accessed 27 June 2019.

European Parliament. 2017. Motion for a European Parliament resolution with recommendations to the Commission on Civil Law Rules on Robotics (2015/2103(INL)). http://www.europarl.europa.eu/doceo/document/A-8-2017-0005_EN.pdf. Accessed 20 March 2019.

Finck, Michèle. 2017. Digital Co-Regulation: Designing a Supranational Legal Framework for the Platform Economy. LSE Law, Society and Economy Working Papers 15/2017. http://dx.doi.org/10.2139/ssrn.2990043. Accessed 4 June 2019.

Fitsilis, Fotios. 2018. Inter-parliamentary cooperation and its administrators. *Perspectives on Federalism* 10(3): E-28-E-55. https://doi.org/10.2478/pof-2018-0030.

Fitsilis, Fotios and Alexandros Koutsogiannis. 2017. Strengthening the Capacity of Parliaments through Development of Parliamentary Research Services. Thirteenth Wroxton Workshop of Parliamentary Scholars and Parliamentarians 29-30 July 2017. http://wroxtonworkshop.org/wp-content/uploads/2017/07/2017-Session-5A-Fitsilis-and-Koutsogiannis.pdf. Accessed 21 March 2019.

Hasselbalch, Gry, 2017. New EU rules for the ethical and legal status of robots and AI. https://dataethics.eu/en/new-eu-rules-ethical-legal-status-robots-ai/. Accessed 20 March 2019.

He, Yujia. 2017. How China is preparing for an AI-powered Future. Wilson Briefs. June 2017. https://www.wilsoncenter.org/sites/default/files/how_china_is_preparing_for_ai_powered_future.pdf. Accessed 21 March 2019.

IPU and IFLA. 2015. Guidelines for parliamentary research services. https://www.ifla.org/publications/node/9759. Accessed 21 March 2013.

JURI. 2017. Report with recommendations to the Commission on Civil Law Rules on Robotics. http://www.europarl.europa.eu/committees/en/juri/subject-files.html?id=20170202CDT01121. Accessed 20 March 2019.

Marx, Axel. 2017. The Public-Private Distinction in Global Governance: How Relevant is it in the Case of Voluntary Sustainability Standards? *The Chinese Journal of Global Governance* 3(1): 1-26. https://doi.org/10.1163/23525207-12340022.

Orozco, David. 2016. The Use of Legal Crowdsourcing ('Lawsourcing') as a Means to Achieve Legal, Regulatory and Policy Objectives. *American Business Law Journal* 53(1): 145-192. https://doi.org/10.1111/ablj.12074.

O'Sullivan, Andrea. 2017. Don't Let Regulators Ruin AI. *MIT Technology Review*, November/December 2017. https://www.technologyreview.com/s/609132/dont-let-regulators-ruin-ai/.

Ponce Del Castillo, Aída. 2017. A law on robotics and artificial intelligence in the EU? Foresight Brief #02-September 2017, European Trade Union Institute, Brussels. https://www.etui.org/content/download/32583/302557/file/Foresight_Brief_02_EN.pdf. Accessed 20 March 2019.

Scherer, Matthew U. 2015. Regulating artificial intelligence systems: Risks, challenges, competencies, and strategies, *Harvard Journal of Law & Technology* 29, no. 2 (Spring):353-400.

Stone, Peter et al. 2016. Artificial Intelligence and life in 2030. One Hundred Year Study on Artificial Intelligence. Stanford University. Report of the 2015 Study Panel. https://ai100.stanford.edu/sites/g/files/sbiybj9861/f/ai_100_report_0831fnl.pdf. Accessed 20 March 2019.

Trubek, David M., and Louise G. Trubek. 2007. New Governance & Legal Regulation: Complementarity, Rivalry, and Transformation. *Columbia Journal of European Law* 13(3): 539-564.

Wachter, Sandra, Brent Mittelstadt, and Luciano Floridi. 2017. Why a Right to Explanation of Automated Decision-Making Does Not Exist in the General Data Protection Regulation. *International Data Privacy Law* 7(2): 76-99. https://doi.org/10.1093/idpl/ipx005.

Zhao, Ben Y. 2017. Response to the question "Should artificial intelligence be regulated?" Quora, 13 August 2017. https://www.quora.com/Should-artificial-intelligence-be-regulated.

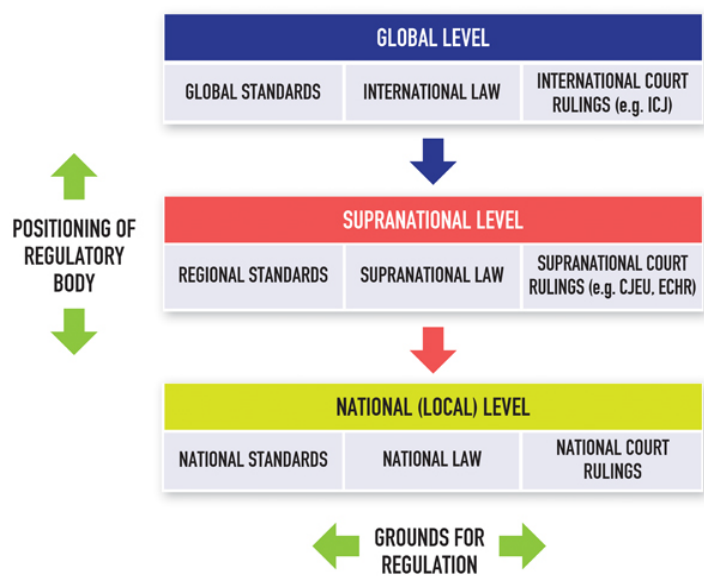

**Fig. 4.1** Hierarchical mode of regulation for advanced algorithms

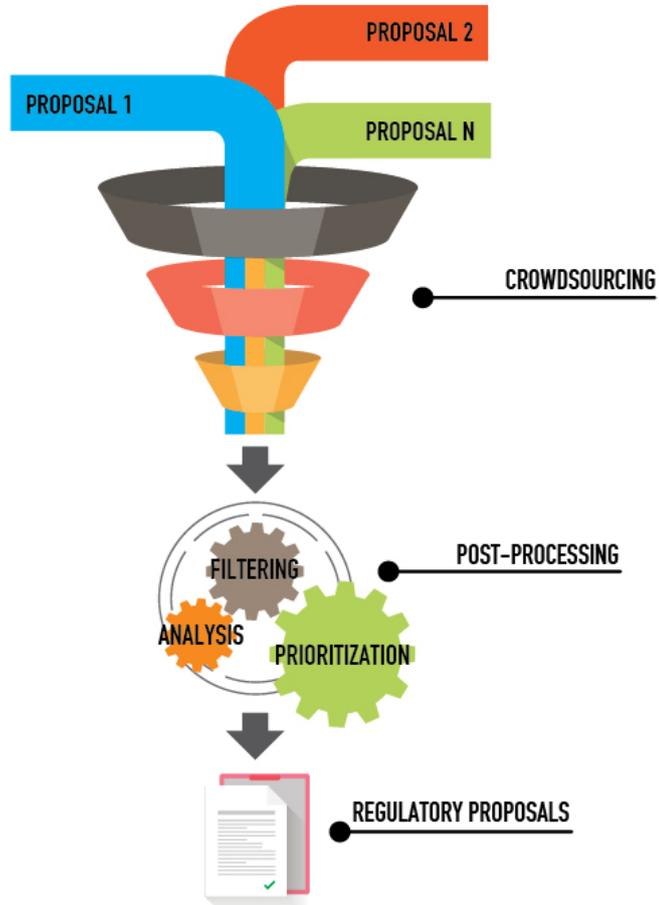

**Fig. 4.2** Principal design of an algorithmic monitor

# 5 Conclusions and Outlook

Fotios Fitsilis

## Abstract

This book offers a new perspective in the discussion around the subject of regulation of advanced algorithms. It presents a number of significant cases where changes in the code and/or the conduct of its developers or operators have been imposed following administrative or judicial decisions. But the existing legal weaponry is often insufficient to directly confront the array of problems related to advanced algorithms. Therefore, the administrative state must begin to employ innovative steps and perhaps aggressive approaches, in order to meet and respond to the challenges highlighted in this book.

Keywords: disinformation, algorithmic malpractice, code of conduct, algorithmic monitor, legislative regulation.

The use of algorithms is not something new, but today's advanced algorithms are indeed far different and more evolved compared to those of the past. In modern digital societies, people conduct confidential and private communications via complex algorithms owned by multinational corporations. That this may result in massive abuse of private data has been demonstrated in the recent Cambridge Analytica scandal, which involved the collection of personally identifiable information of millions of Facebook users, in order to influence US voters during a run-up to a general election. Those revelations led to widespread discussions on the regulation of algorithms and secured Facebook's CEO a testimony to Senate committees as well as to the EUP about the company's data collection practices.[1] As a result, some scholars support the case that important decisions should always remain in the hands of humans, thus eliminating fears of such decisions being made in a non-transparent way and without any accountability or recourse.[2] 
  Democracy itself may very well have died back in its birthplace in Athens, Greece, some 2500 years ago. Contemporary 'democratic' systems are a mere evolution of the original notion. One simply needs to ask the question whether our western democratic systems are in the position to sufficiently govern societies of billions. Well, maybe they are not. In this sense, any recommendation towards an alternative approach in governance is more than welcome.[3] Under these circumstances, the role of advanced algorithms in contemporary and future democracies deserves more analysis, a giant task that is out of the scope of this book.[4]

---

[1] It is these revelations that may have led Facebook's CEO to call for more online regulation and a 'more active role for governments and regulators'; see Zuckerberg (2019).

[2] See, e.g., Coglianese and Lehr (2017), p. 26, in combination with Van Loo (2017), p. 1274.

Yet, in this regard, it is worth mentioning that the issue of disinformation raises broader concerns of societal harm. In the worst-case scenario, the impact of disinformation campaigns can affect entire societies, be that through interference in election, or misinformation about foods, or medicines. The stakes involved in accurately identifying disinformation are high because it often affects free exchange of ideas and information, the core of political discourse, a point that should be of particular concern to parliaments. Notably, in an effort to understand the spread and impact of disinformation as well as ensuring the transparency of the 2019 European Elections, digital platforms including Facebook, Google and Twitter signed up to a voluntary EU Code of Practice on disinformation.[5] Moreover, Google laid out a process to curb fake news as malicious actors have attempted to harm or deceive on-line search users through a wide range of actions.[6] Not to be underestimated in terms of its regulatory impact is also the relevant announcement by the G20 Trade Ministers and Digital Economy Ministers that 'AI actors should respect the rule of law, human rights and democratic value, (…)'.[7]

Since the invasion of new technologies into our lives seems inevitable and at the same time very profitable for a series of private actors, it is evident, more than ever, that the administrative state needs to adjust to this new digital environment. This book has therefore offered some new perspectives in the discussion around the regulation of advanced algorithms by presenting a number of significant cases where changes in software have been imposed following administrative or judicial decisions. These cases have been classified and analyzed in order to provide insights to the inner mechanisms of algorithmic regulation.

---

[3] Hence, the significance of the contribution by Runciman (2018).

[4] Nevertheless, this can be linked to our proposal to work on underlying legal values and principles rather than to produce scores of regulation.

[5] See European Commission (2019); the Code is an initiative of the European Commission (2018).

[6] These actions include for instance tricking online systems in order to promote their own content (via a set of practices referred to as 'spam'), propagating malware, and engaging in illegal acts online, Google (2019).

[7] See G20 (2019), p. 11; this is the first time the countries that represent world's top economies agreed upon a set of principles for 'responsible stewardship of trustworthy AI', which may serve as guidelines for national and supranational regulatory policies.

| Topic | Legislative regulation | Judicial regulation | Regulatory acts | Legal action | Legal basis | Regulation on algorithm | Regulation on controller |
|---|---|---|---|---|---|---|---|
| **MS Media Player and Explorer case** | | ● | | *US v. Microsoft* 231 F. Supp. 2d 144 | 15 USC §§1-2 15 USC§16(b)-(h) | | ● |
| | | | | *Microsoft v. EC* T-201/04 | Art. 81&82 ECT 91/250/EEC | ● | ● |
| **Emission case** | ● | ● | (EU) P8_TA-PROV(2018)0179 | *US v. VW* 16-CR-20394 | 18USC§371 18USC§1512(c) 18USC§542 | | ● |
| **Ad-block case** | | ● | | *Axel Springer AG v. Eyeo* | §4 Nr. 4 UWG & §4a UWG | | ● |
| **Block online content** | ● | ● | (EU)2016/679 | *Google Spain SL & Google Inc. v. AEPD&González* | 95/46/EC 2000/C 364/01 | ● | ● |
| **Sharing economy a. AirBnB case** | ● | ● | *Minpaku Shinpou* (JP) *Nomos* 4472/2017 (GR) *Loi* n° 2016-1321 (FR) | *Airbnb France v. City of Paris* 18-40.043 | Art. L.324-2-1 & L.324-1-1 Tourism Code; Art. L. 631-7 et seq. Construction and Housing Code (FR) | ● | ● |
| **Sharing economy b. Uber case** | | ● | *Loi* n° 2014-1104 (FR) *Nomos* 4530/2018 (GR) | *Elite Taxi v. Uber Systems Spain* | 2006/123/EC | | ● |
| | | | | *Philadelphia Taxi Assoc. Inc. v. Uber* | 15 USC § 2 15 USC § 15 | | |
| | | | | *Sorvano v. Uber* | 15 USC § 1125(a) CBPC §17200 et seq. | | |
| | | | | *Razak v. Uber* | 29 USC § 201 et seq. | | |
| **Algorith-mic trading** | ● | ● | 2014/65/EU US Public Law No: 111-203 | *Coscia v. US* 17-1099 | 7 USC §6c(a)(5) 7 USC §6c(a)(5)(C) 7 USC §9 18 USC§1348 | ● | ● |
| **GDPR** | ● | | (EU)2016/679 | | | ● | ● |

**Table 5.1** Overview of analyzed topics on advanced algorithms

Table 5.1 presents a wide overview of the topics of regulation this book refers to. The 'type mode' of regulation, which consists of the legislative and judicial regulatory approach, is displayed, along with the relevant regulatory acts, such as directives, regulations, laws and decrees, or legal action that may apply to each case. In case a higher instance has confirmed a decision of a lower court, the indicated case(s) may deviate from the one(s) discussed under the relevant topic.[8] Finally, the regulatory outcome is indicated, i.e. whether regulation has been imposed on the environment or the controller (indirect regulation), or on the algorithm itself (direct regulation). We speak here of the nature of regulation ('natural mode').

Not all cases that involve algorithms are cases of direct regulation, in terms of imposing (or aiming at) changes directly to the algorithm,[9] such as in the Microsoft Windows Media Player case, which led to the creation of 'Windows XP N', a Windows operating system without the media player[10]. Some cases may relate to pure labor issues,[11] or competition[12] or restrictive measures to the developer / controller of an algorithm.[13] There, regulation may be regarded as 'indirect', i.e. when conditions are imposed on the surrounding environment of the algorithm or its controller. It is difficult to tell, whether classification according to the natural mode provides any hints regarding the gravity of intervention. This may be the case, to the extent regulation affects the viability of the underlying business model the algorithm serves or its principal functionality. Given the lack of appropriate general regulatory principles on algorithmic regulation, or a sector specific legal framework, 'traditional' legal instruments, such as competition law, have been frequently used for algorithmic regulation. Classification in regulatory types may also include the timing of the intervention.[14]

Hence, overall we classify regulation into five modes and their respective subclasses, according to the type (legislative or judicial), nature (direct or indirect), timing (*ex-ante* or *ex-post*) and form of the intervention (command-and-control, self-regulation or co-regulation). These are embedded into a vertical dimension, the 'hierarchical mode' (global/supranational/national). With the form of interven-tion already presented elsewhere[15] and the 'hierarchical mode' being intrinsic to any form of regulation, we hereby underline the importance of the remaining three modes, i.e. 'type', 'nature' and 'timing', for providing added-value to the surrounding discussion. Fig. 5.1 visualizes these regulatory modes and depicts their attributes and interrelations.

---

[8] See for instance Section 3.6: Algorithmic trading regulation.

[9] The notion of change in the architecture of the code is described by Lessig (1999), pp. 505-506.

[10] See EC Decision of 24 March 2004, case COMP/C-3/37.392.

[11] *Razak v. Uber Technologies Inc.*, 2:16-cv-00573.

[12] *Philadelphia Taxi Assoc. Inc. v. Uber*, 17-1871.

[13] The MiFID II / MiFIR framework and the GDPR offer several such provisions.

[14] Usually, the 'type mode' is equivalent to the 'timing mode'; however, this may not always be the case, as shown in Chapter 1.

The EU and US cases, which involve anti-competitive bundling of Microsoft's internet browser and media player with the underlying operating systems, have been paramount in shaping the computer and software industry. Regulation here was achieved by judicial decisions rather than using a specific legal framework. However, in both legal orders, anti-competitive provisions of Art. 101 and 102 TFEU[16] and Sections 1 and 2 of the Sherman Antitrust Act have been used as a general legal basis. Interestingly, the settlement achieved by Microsoft in the US did not require the company to alter the code by removing IE from its operating system. An EC decision against Microsoft, affirmed in its essence by the Court of First Instance, forced Microsoft to create a different version of its operating system without the Windows Media Player.

In the VW diesel emissions case, US investigators found out that the company had used a defeat device in the form of an illegal algorithmic switch, which sensed the operating conditions of the vehicle and adjusted gas emissions accordingly. Within the context of the present book, the basic US case[17] has been discussed as well as the relevant settlements as VW pled guilty to a series of criminal and civil charges. At the same time, the legal framework in the EU at the time did not allow for an analogous prosecution of VW on a European level. What is of particular interest is the fact that VW, both in the US and the EU, has not been explicitly urged to immediately cease using the mentioned defeat device. Only after talks with the EC, VW committed to resolve the issue for the European consumers via a software update.

The inter-relations between online privacy, digital marketing and fair competition have been discussed before German courts in an interesting case of legitimacy in using ad-blocking software. Following legal action from Axel Springer AG, a private media company, BGH, the German Federal Court, declared the operation of AdBlock Plus lawful. According to the Court, any (financial) damage was not caused by the software provider but by the users who downloaded the ad blocker. In addition, it was noted that media companies have always the possibility to block users that use ad-blocking software.

The boundaries of the personal right to block online controversial content, which is widely known as the 'right to be forgotten', have been set by the CJEU in its landmark judgement of the case *Google Spain SL and Google Inc. v. AEPD*

---

[15] In a Communication by the European Commission (2016, p. 5), self-regulation and co-regulation mechanisms are suggested as command-and-control alternatives for regulating platform economy.

[16] Formerly known as Art. 81 and 82 ECT, respectively.

[17] *US v. Volkswagen*, 16-CR-20394.

*and González*. The Court's ruling, which applies to search engines with a branch or subsidiary in the EU, provides data subjects with the right to have their personal information removed from relevant searches on the basis of their name, even if this information appears on the indexed pages in a lawful way. Search engine operators have to comply with such requests while keeping a number of factors in balance such as their own economic viability and the right of access to information of the general public. The 'right to be forgotten' has been incorporated in Art. 17 Regulation (EU) 2016/679 that has significant implications both for system operators/controllers and the utilized software/algorithms.

The sharing economy, as expressed through the Airbnb and Uber cases, took traditional industries by surprise. In numerous countries worldwide, the latter are seeking state or judicial protection in order to avoid collapse. Both Airbnb and Uber rely on algorithms that match supply with demand with the companies cashing a small provision when the arranged service has been provided. In the home-sharing cases, such as the case of Airbnb,[18] local regulators, e.g. city administration, or the central government, are imposing strict limitations to the landlords. On the other hand, in several cases, platforms are asked not to list any offers that have not been declared with the competent authorities and do not bear an official registration number. In the Uber case, most resistance comes from the traditionally strong taxi-service syndicates. A couple of US lawsuits[19] have been handpicked in order to present the state of play in the car-sharing business. Similarly, to a different case concerning ad-blocking in Germany, *Axel Springer AG v. Eyeo*, competition laws have been invoked in the first lawsuit to protect the existing status quo in the industry. With argumentation analogous, to a certain extent, to the BGH ruling, the US Court of Appeals rejected the plaintiff's arguments by stating that the taxi association had no right to exclude competitors from the market. This implies that taxi companies would need to become more competitive in order to continue to be viable in this business.[20] The second pending case involves allegations of algorithmic manipulation by Uber that contain parallels with the VW emissions case.

Algorithmic trading and HFAT have been at the center of regulatory policies and judicial decisions as they bear potential for serious market distortion. In recent years, the European Union has pioneered financial markets regulation through MiFID and the MiFID II / MiFIR framework. In the US, the regulatory framework involves the Dodd–Frank Wall Street Reform and Consumer Protection Act. Based on the latter, unlawful HFAT practices have been prosecuted and the outcome of a significant case in the US has been discussed. In this complex field, regulations are imposed to both algorithms, which are to operate within certain limits, as well as to the controllers and trading venues, such as strict reporting requirements and rules on the admission of financial instruments to trading.

---

[18] Regulation applies to other online platforms as well, such as Homestay, Couchsurfing, Home Exchange, Wimdu, Bedycasa and Culture Go Go.

[19] The two lawsuits are *Philadelphia Taxi Assoc. Inc. v. Uber*, 17-1871 and *Sorvano v. Uber*, 2:17-cv-02550-DMG-JEM.

[20] The analogy with the BGH ruling lies in the fact that the German Federal Court stated that there is no general market obstruction and that the plaintiff seeking judicial protection must become active in order to 'protect' its business. What is remarkable here is the fact that competition law has been used for the judicial reasoning in different legal orders both in the EU (in Union and member state level) and the US.

Following rapid developments in technology and learning from cases of the past, some of which have been discussed herein, the EU went a decisive step further towards an ex-ante regulation of personal data protection. This action by the EU institutions took the form of a Regulation, the GDPR. The GDPR contains some relevant legal principles from pre-existing data protection law, enriches them and incorporates new elements for enhanced personal data protection. In the present context, the concepts of data protection *by design* and *by default* have been analyzed and some special parameters of data processing, such as *pseudonymization* and *data minimization* have been discussed.

The present book examined therefore a series of major cases where advanced algorithms came to play. One of the issues that has been investigated dealt with the research question as to whether existing legal instruments are sufficient to confront an array of problems related to advanced algorithms. Legislative (or ex-ante) regulation in a field that leaps forward proves indeed difficult. Analysis shows that despite ongoing algorithmisation of administrative decisions and private operations, only the EU and possibly China (although the level of analysis herein is not sufficient) are proceeding with a general legislative framework to regulate algorithmic conduct. The EU is expressing its will to be a front-runner in establishing a general legislative framework for algorithmic regulation. In this regard, the GDPR and MiFID II, in data privacy and algorithmic trading regulation respectively, offer clear indications of a more aggressive centralized approach compared to the US. On the other hand, the US is more decisive when it comes to legal action, with its prosecutors even charging perpetrators with criminal charges, such as in the VW case, which itself amounted to commercial fraud and intended misselling of a product.

Nevertheless, even a firm regulatory framework cannot foresee any biases in algorithms or reveal algorithmic malpractices. This can only be achieved through in-depth investigations and thorough analysis by appropriate and highly specialized bodies. In this regard, the establishment of a relevant agency for algorithmic regulation and control is proposed. According to the approach presented herein, however, it is not necessary to establish a new institution or agency from scratch. In order to save valuable time and resources,[21] the competencies of an existing agency could be expanded in order to cover the aforementioned issues. Hence, given the nature of advanced algorithms, such as complexity and early stage of development, this contribution to the field considers ENISA, a European agency that focuses on network and information security, or an appropriate spin-off, to be an interesting candidate for this role.

---

[21] See relevant discussion on the EU level on the establishment of an agency with market surveillance powers to oversee road transport. Here the main arguments against it focused on the high cost for its implementation.

National or supranational parliaments could also play an important role in algorithmic regulation, particularly in the course of their *oversight function*. Their role could be further strengthened by establishing special mechanisms for following up implementation of laws and by-laws, a process that is called post-legislative scrutiny. For this, it is necessary to increase their administrative and scientific capacity, through for example further development of their parliamentary research services. In view of the rise of advanced algorithms and the overall significance of representative institutions in democratic societies, a dedicated study should be conducted in order to analyze the parameters and conditions under which parliaments could contribute in the field of algorithmic regulation.

A solid legal framework is a non-plus-ultra to achieve quick and well-founded legal decisions in the many legal disputes that are going to take place in the time yet to come. As several cases where algorithms come to play have been legally assessed, scrutiny illustrates that the existing legal weaponry is still not sufficient to directly and efficiently regulate advanced algorithms. Instead, antitrust and competition laws have been utilized in many of the discussed cases. In a rapidly developing field with ever shortening life cycles of advanced algorithms and of the related software products or platforms, yearlong investigations in such disputes will practically result in a denial of justice. Specialized, rather than general,[22] legislation, such as EU's GDPR/MiFID II, cooperation between state institutions around the globe, such as in the Microsoft cases, and the evolution of dedicated agencies, will be necessary to spot problematic algorithmic cases and efficiently tackle related issues even before they arise. An alternative approach would be an attempt to develop a rigid regime of legal values, along with a set of related rights that apply while designing, implementing and operating advanced algorithms.[23] The definition of those values and rights should be affected on the supranational level, such as in the form of a convention or resolution, rather than at the national level and it is a prerequisite for them to be applied by judges in an efficient manner.

Overall, in the wide regime of advanced algorithms, technology moves faster than governments can address its effects and a clear governmental regulatory pattern does not seem to exist. But even in cases of regulation through administrative decisions, the competitive nature of the research field and the projections of its future market value may result in a situation where those decisions are always contested in front of the competent courts and even up to the highest instance. As a result, judicial (or ex-post) regulation is the rule. This is not expected to change in future algorithmic cases, even with further development of dedicated specialized legislation. Hence, the –constitutional– right to a speedy trial could be guaranteed by the definition of general legal values applicable to algorithmic cases, as mentioned previously.

---

[22] E.g. competition or labor law

[23] The newly forged 'right to be forgotten' may constitute such a right.

Algorithmic malpractices, such as concerns regarding discrimination and breach of privacy, just to name a few, are subject of a heated debate. While regulators may ensure that data subjects are protected against private sector offenders, a legitimate question to be asked is whether these are also sufficiently protected from algorithmic malpractices by state organs. Differently formulated, and in the absence of tailored legal values, how can it be ensured that the basic principles of public law, such as non-discrimination, accountability, transparency, are not being violated by the state itself? The outcome of our on-going investigation to this question is of particular significance, as it touches upon fundamental individual and collective rights that are essential for the functionality of modern democratic societies. The development of a code of conduct for public sector agencies, the establishment of an 'ethics advocate' or a mere 'trusted third party', who represents an independent and trustworthy expert[24] within critical public service units and continuous professional training on privacy, legitimacy and democratic values, all constitute efficient tools to counterbalance informational superiority of the State against private data subjects and to ensure the legality of its actions.

The main parameters that have been analyzed for each of the described cases were location (EU, US or elsewhere), administrative or judicial reasoning and legal basis. As a next step, the geographical criterion could be expanded to cover more cases from other continents. A general finding of the book is that the European Union is quick in taking legislative action, whereas the US is quick in taking legal action. Further cases of algorithmic regulation are necessary to be studied to support this claim, however. Additional research is also necessary to determine the operational framework of a mechanism to screen cases where algorithms come to play and potentially need regulation. This so-called 'algorithmic monitor', which could also involve crowdsourcing, is considered to offer essential input to the work of a regulatory body. The form and operation of such an apparatus requires significant study itself.

Concluding, despite the aforementioned concerns that come with the use of algorithms, governments should not be hesitant to invest in their immense possibilities to ameliorate the administrative state. Still, such technologies are not ripe enough and we should use the time for the planning of regulatory principles and law-making. Scientific foresight and forward-thinking legal assessment should be widely employed in order to determine and regulate the effects of advanced algorithms in and for future societies.

---

[24] Alternatively, a body of experts could be regarded, depending on the complexity and significance of the administrative decisions to be made.

# References

Coglianese, Cary, and David Lehr. 2017. Regulating by Robot: Administrative Decision Making in the Machine-Learning Era. Research Paper No. 17-8. Institute for Law and Economics. University of Pennsylvania. https://www.law.upenn.edu/live/files/6329-coglianese-and-lehr-regulating-by-robot-penn-ile. Accessed 18 March 2019.

European Commission. 2016. Communication on Online Platforms and the Digital Single Market.

Opportunities and Challenges for Europe. COM (2016) 288 final. https://eur-lex.europa.eu/legal-content/EN/TXT/PDF/?uri=CELEX:52016DC0288&from=EN. Accessed 27 June 2019.

European Commission. 2018. Code of Practice on Disinformation. https://ec.europa.eu/newsroom/dae/document.cfm?doc_id=54454. Accessed 6 June 2019.

European Commission. 2019. Code of Practice against disinformation: Commission calls on signatories to intensify their efforts. Press release, January 29, 2019. http://europa.eu/rapid/press-release_IP-19-746_en.htm. Accessed 31 May 2019.

Google. 2019. How Google Fights Disinformation: 9-16. https://storage.googleapis.com/gweb-uniblog-publish-prod/documents/How_Google_Fights_Disinformation.pdf. Accessed 31 May 2019.

G20. 2019. G20 Ministerial Statement on Trade and Digital Economy. June, 8-9. Tsukuba, Japan. https://g20trade-digital.go.jp/dl/Ministerial_Statement_on_Trade_and_Digital_Economy.pdf. Accessed 25 June 2019.

Lessig, Lawrence. 1999. The law of the horse: What cyberlaw might teach. *Harvard Law Review* 113.2: 501-546.

Runciman, David. 2018. *How democracies end*. London: Profile Books.

Van Loo, Rory. 2017. Rise of the digital regulator. *Duke Law Journal* 66:1267-1329.

Zuckerberg, Mark. 2019. The Internet needs new rules. Let's start in these four areas. *The Washington Post,* March 30, 2019. https://www.washingtonpost.com/opinions/mark-zuckerberg-the-internet-needs-new-rules-lets-start-in-these-four-areas/2019/03/29/9e6f0504-521a-11e9-a3f7-78b7525a8d5f_story.html?noredirect=on&utm_term=.751b6e9e19e7. Accessed 31 May 2019.

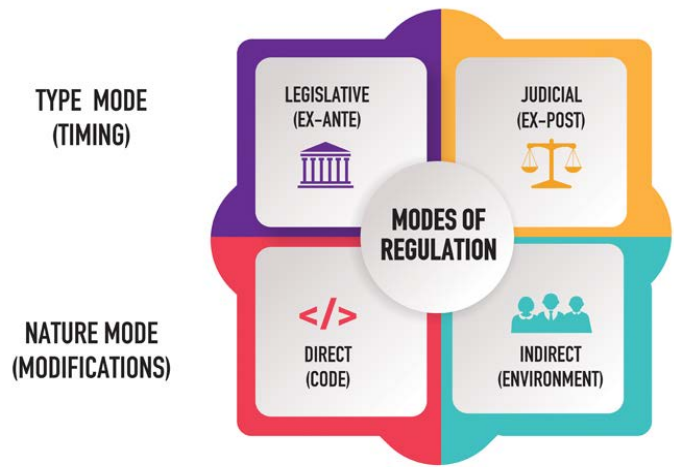

**Fig. 5.1** Advanced classification of regulatory measures: natural, type and timing modes

# Appendix

## Tables

| Case | Identifier | Document type | Year |
|------|-----------|---------------|------|
| Airbnb France v. City of Paris | 18-40.043 | Judgement of the Court of Cassation (Third Civil Chamber) | 2019 |
| Asociación Profesional Elite Taxi v. Uber Systems Spain, SL | C- 434/15 | Judgment of the Court (Grand Chamber) | 2017 |
| Axel Springer AG v. Eyeo | I ZR 154/16 BGH | Press announcement, BGH | 2018 |
| City of Paris v. Airbnb France and Airbnb Ireland | 18-54.632 | Judgement, TGI | 2019 |
| City of Paris v. Ms. Claire G.F. | 17-26.158 | Judgement of the Court of Cassation (Third Civil Chamber) | 2018 |
| Coscia v. United States | 17-1099 | Docket for 17-1099 Supreme Court, US | 2018 |
| Google Spain SL and Google Inc. v. AEPD and Mario Costeja González | C-131/12 | Judgment of the Court (Grand Chamber) | 2014 |
| Microsoft v. European Commission | T-201/04 | Judgment, Court of First Instance (Grand Chamber) | 2007 |
| M.L. and W.W. v. Germany | ECHR 554 | Judgment (fifth section) | 2018 |
| New York v. Microsoft Corp. | F. Supp. 2d 132 (D.D.C. 2002) | Memorandum Opinion | 2002 |
| Philadelphia Taxi Association, Inc. v. Uber Technologies | 17-1871 | Opinion, US Court of Appeals for the Third Circuit | 2017 |
| Razak v. Uber Technologies Inc. | 2:16-cv-00573 | Memorandum Re: Defendants' Motion for Summary Judgment | 2018 |
| Scarlet Extended SA v. SABAM | C-70/10 | Court Judgment (Third Chamber) | 2011 |
| Sorvano v. Uber | 2:17-cv-02550-DMG-JEM | Class action complaint | 2017 |
| US v. Coscia | 14-cr-00551 | Memorandum Opinion | 2014 |
| US v. Microsoft Corp. | 231 F. Supp. 2d 144 (D.D.C.2002) | Memorandum Opinion | 2002 |
| US v. Sarao | 1:15-cr-00075 | Plea Agreement | 2016 |
| US v. Volkswagen | 16-CR-20394 | Plea Agreement | 2017 |

**Table A** Overview of court cases

| Institution/Organization | Identifier | Title | Year |
|---|---|---|---|
| Article 29 Working Party (WP29) | 17/EN - WP251rev.01 | Guidelines on Automated individual decision-making and Profiling for the purposes of Regulation 2016/679 | 2018 |
| *Assemblée Nationale*, FR | *Loi* (Law) n° 2014-1104 | Law relating to taxis and chauffeur-driven transport | 2014 |
| *Assemblée Nationale*, FR | *Loi* (Law) n° 2016-1321 | Law for a Digital Republic | 2016 |
| *Assemblée Nationale*, FR | *Loi* (Law) n° 2016-1918 | Budget Law | 2016 |
| *Assemblée Nationale*, FR | *Loi* (Law) n° 2018-898 | Anti-Fraud Act | 2018 |
| Congress, US | 42 USC 7401–7626 | Clean Air Act | 1963 |
| Congress, US | Public Law No: 111-203 | Dodd–Frank Wall Street Reform and Consumer Protection Act | 2010 |
| Congress, US | S.2123 - 114th Congress (2015-2016) | Sentencing Reform and Corrections Act (pending) | 2015 |
| Council of Europe | Resolution 2051 (2015) | Drones and targeted killings: the need to uphold human rights and international law | 2015 |
| Environmental Protection Agency, US | SEP 18 2015, NOV – 2 2015 | Notice of Violation (1$^{st}$ & 2$^{nd}$) | 2015 |
| European Parliament | 2015/2103(INL) | Report with recommendations to the Commission on Civil Law Rules on Robotics | 2017 |
| European Parliament | 2015/3037(RSO) | Committee of Inquiry into Emission Measurements in the Automotive Sector | 2015 |
| European Parliament | 2016/2215(INI) | Committee report with findings and conclusions | 2016 |
| European Parliament | EPRS, 13 July 2017 | Public consultation on robotics and AI - First results of public consultation | 2017 |
| European Union | 2000/C364/01 | Charter of Fundamental Rights of the European union | 2000 |
| European Union | 2000/31/EC | Directive on electronic commerce | 2000 |
| European Union | 2001/29/EC | Copyright Directive | 2001 |
| European Union | 2002/58/EC | Directive on privacy and electronic communications | 2002 |
| European Union | 2004/48/EC | Directive on the enforcement of intellectual property rights | 2004 |
| European Union | 2006/123/EC | Services in the internal market | 2006 |
| European Union | 2007/46/EC | Approval of motor vehicles | 2007 |

| | | and their trailers, and of systems, components and separate technical units intended for such vehicles | |
| European Union | 2014/65/EU | Markets in Financial Instruments Directive II | 2014 |
| European Union | 2016/679/EU | General Data Protection Regulation (GDPR) | 2016 |
| European Union | 2015/2120/EU | Regulation on roaming on public mobile communications networks within the Union | 2015 |
| European Union | 2018/858/EU | Vehicle type approval framework regulation | 2018 |
| European Union | COM/2016/0593 final - 2016/0280 (COD) | Proposal for a Directive on copyright in the Digital Single Market | 2016 |
| Hellenic Data Protection Authority, GR | Decision 1/2017 | Notification for processing of personal data for electronic ticket application | 2017 |
| Hellenic Data Protection Authority, GR | Decision 4/2017 | Processing of personal data in the framework of the unified automatic fare collection system | 2017 |
| Hellenic Parliament, GR | *Nomos* (Law) 4472/2017 | Art. 83 and 84 regulate income taxation and other issues relating to house sharing, respectively | 2017 |
| Hellenic Parliament, GR | *Nomos* (Law) 4530/2018 | Regulation of transport issues and other provisions | 2018 |
| House of Commons, UK | Minutes of Proceedings, 2 May 2018 | Breast Cancer Screening | 2018 |
| Information Commissioners' Office, UK | - | Report on the Consultation, GDPR consent guidance | 2017 |
| International Court of Justice | - | Statute | 1945 |
| Inter-Parliamentary Union & International Federation of Library Associations and Institutions | ISBN 78-92-9142-630-0 | Guidelines for PaRS | 2015 |
| *Juutaku shukuhaku jigyohō* 'Minpaku Shinpou', JP | Act No. 65 | Private House Lodging Business Act | 2017 |
| *La Mairie de Paris* (Paris City Hall), FR | 2017 DLH 128 | Obligation to register a property according to Art. L314-1-1 of the Tourism Code | 2017 |

**Table B.** Overview of presented decisions, directives, laws, regulations, resolutions and reports

# Index